\documentclass[11pt]{article}
\pdfoutput=1
\usepackage{jheppub}
\usepackage[T1]{fontenc}
\usepackage[utf8]{inputenc}
\usepackage{cancel,tensor,enumitem,fix-cm}
\usepackage{graphicx}
\usepackage[english]{babel}
\usepackage{amsmath,amssymb}
\usepackage{mathabx}
\usepackage{textcomp}
\usepackage{multirow}
\usepackage{booktabs}
\usepackage{tikz}
\usepackage{overpic}
\usepackage{slashed}

\newcommand{\bea}{\begin{eqnarray}}
\newcommand{\eea}{\end{eqnarray}}
\newcommand{\be}{\begin{equation}}
\newcommand{\ee}{\end{equation}}

\def\({\left(}
\def\){\right)}

\def \a {\alpha}

\def \k {\kappa}

\def \l {\lambda}
\def \m {\mu}
\def \n {\nu}

\def \t {\tau}

\def \D {\Delta}

\def\nn {\nonumber}

\subheader{\begin{flushright}
\end{flushright}}

\title{Inverse Anisotropic Catalysis in Holographic QCD}

\author[\dag]{Umut G\"ursoy,}
\author[\dag]{Matti J\"arvinen,}
\author[\dag]{Govert Nijs}
\author[\ddag]{and Juan F. Pedraza}
\affiliation[\dag]{Institute for Theoretical Physics and Center for Extreme Matter and Emergent Phenomena, Utrecht University, 3584 CE Utrecht, The Netherlands}
\affiliation[\ddag]{Institute for Theoretical Physics, University of Amsterdam, 1090 GL Amsterdam, The Netherlands}
\emailAdd{u.gursoy@uu.nl}
\emailAdd{m.o.jarvinen@uu.nl}
\emailAdd{g.h.nijs@uu.nl}
\emailAdd{jpedraza@uva.nl}

\abstract{
We investigate the effects of anisotropy on the chiral condensate in a holographic model of QCD with a fully backreacted quark sector at vanishing chemical potential. The high temperature deconfined phase is a neutral and anisotropic plasma showing different pressure gradients along different spatial directions, similar to the state produced in noncentral heavy-ion collisions.
We find that the chiral transition occurs at a lower temperature in the presence of anisotropy.
Equivalently, we find that anisotropy acts destructively on the chiral condensate near the transition temperature. These are precisely the same footprints as the ``inverse magnetic catalysis''  i.e. the destruction of the condensate with increasing magnetic field observed earlier on the lattice, in effective field theory models and in holography. Based on our findings we suggest, in accordance with the conjecture of \cite{Giataganas:2017koz}, that the cause for the inverse magnetic catalysis may be the anisotropy caused by the presence of the magnetic field instead of the charge dynamics created by it.
We conclude that the weakening of the chiral condensate due to anisotropy is more general than that due to a magnetic field and we coin the former ``inverse anisotropic catalysis''.
Finally, we observe that any amount of anisotropy changes the IR physics substantially: the geometry is $\text{AdS}_4 \times \mathbb{R}$ up to small corrections, confinement is present only up to a certain scale, and the particles acquire finite widths.}

\begin{document}
\maketitle
\flushbottom

\section{Introduction}

Understanding all corners of the phase diagram of
quantum chromodynamics (QCD) is a major focus of current research.
Besides theoretical curiosity, studying QCD matter in extreme conditions
is crucial in many physical situations ranging from the ultra-relativistic heavy-ion collision experiments at RHIC and LHC, to the core of neutron stars and magnetars, and to early cosmology \cite{Duncan:1992hi,Vachaspati:1991nm, Tashiro:2012mf,Semikoz:2007ti,Semikoz:2009ye,Semikoz:2012ka, Kahniashvili:2012uj, Dvornikov:2013bca}.

According to our current understanding, colliding heavy ions
create a strongly-coupled deconfined plasma state known as the quark-gluon plasma (QGP) that behaves almost as a perfect fluid \cite{Gyulassy:2004zy,Muller:2006ee,Tannenbaum:2006ch,Shuryak:2006se,dEnterria:2006mtd,Muller:2007rs,Zajc:2007ey}. In the event of off-central, i.e. with nonvanishing impact parameter, collisions the plasma is highly anisotropic\footnote{Anisotropy is present even in the central collisions, thanks to fluctuations in the initial shape of the participant nuclei. This can be inferred from the fact that the elliptic flow parameter $v_2$ is typically non-vanishing even at vanishing centrality, see e.g. \cite{Aamodt:2010pa, Adam:2016izf, Ackermann:2000tr}.} and is typically created in the presence of a strong magnetic field which can reach up to $eB/m_\pi^2 \sim 5-10$ \cite{Skokov:2009qp,Tuchin:2010vs,Voronyuk:2011jd,Deng:2012pc,Tuchin:2013ie,McLerran:2013hla,Gursoy:2014aka}.

The interplay between strong magnetic fields, strong interactions and finite temperature has been studied extensively in the literature, and is known to lead to rich phenomenology, see \cite{Kharzeev:2012ph, Kharzeev:2013jha, Miransky:2015ava} for reviews. One of the most surprising effects in this context is the phenomenon known as {\em Inverse Magnetic Catalysis} (IMC)\@. IMC refers to the observation that, at temperatures of the order of $\sim150$ MeV, the presence of a strong external magnetic field has a destructive effect on the chiral condensate $\langle\bar{q}q\rangle$ \cite{Bali:2011qj,Bali:2011uf,Bali:2012zg,DElia:2012ems}. This phenomenon was first observed on the lattice, and cannot be explained by standard perturbative calculations. In fact, perturbative QCD predicts the exact opposite effect, dubbed as {\em Magnetic Catalysis} (MC) \cite{Gusynin:1994re,Gusynin:1994xp,Gusynin:1994va}. The intuition behind the Magnetic Catalysis is that, in the presence of a strong magnetic field, charged particles freeze in their lowest Landau level, effectively reducing the dimensionality to $(1+1)$. Since the IR  dynamics  in  gauge  theories  in  lower  dimensions is much stronger in comparison to their higher dimensional counterparts, this leads to a strengthening of the condensate and  catalysis  of  chiral  symmetry  breaking \cite{Miransky:2015ava}. The lattice results of \cite{Bali:2011qj,Bali:2011uf,Bali:2012zg,DElia:2012ems} on the other hand indicate that the inverse effect, i.e. weakening of the condensate  arises again from the strongly coupled dynamics around the deconfinement temperature at stronger magnetic fields.

The exact mechanism that leads to IMC remains elusive today. One compelling idea  \cite{Bruckmann:2013oba,Bruckmann:2013ufa} based on lattice calculations,
is that IMC arises due to competition between the ``valence'' and the ``sea'' quarks in the quark propagator\footnote{For simplicity, we consider only one fermion
flavor with  mass $m$, but the idea applies more generally.}:
\be\label{qqint}
\langle\bar{q}q\rangle_B=\frac{1}{\mathcal{Z}(B)}\int \mathcal{D}A_\mu^a e^{-S_g}\det (\slashed{D}(B)+m)\text{Tr}\,(\slashed{D}(B)+m)^{-1}\,,
\ee
where $\mathcal{Z}(B)$ is the path integral without the propagator, the trace and the determinant are taken over the spin and the momentum space and $\slashed{D}(B)$ includes coupling of fermions both to the external magnetic field and to the gluons $A_\mu^a$. For a magnetic field in the $x^3$-direction
\be\label{DB}
\slashed{D}(B) = \gamma^{\mu}\left( \partial_\mu + A^a_\mu T^a + e A^B_\mu\right), \qquad A^B_\mu = (0, Bx^2/2, -Bx^1/2,0)\,.
\ee
The ``valence'' contribution arises from the quark operators inside the path integral (\ref{qqint}) i.e. from the trace. The effect of $B$ through this contribution always tend to catalyze the condensate simply because $B$ increases the spectral density of the zero energy modes of the Dirac operator. The ``sea'' contribution, on the other hand, comes from the determinant that describes fluctuations around the gluon path integral. The $B$ dependence of this contribution suppresses the condensate around the deconfinement temperature \cite{Bruckmann:2013oba,Bruckmann:2013ufa}. Another idea is based on a competition between the total magnetic field dependence of the quark propagator and of the QCD coupling constant when RG scale is taken at $B$ \cite{Miransky:2015ava}.

In this paper, we put forward and test an alternative idea:  inverse magnetic catalysis results from the {\em anisotropy} in the system caused by the presence of the external magnetic field or by other means. Imagine, as a result of some mechanism, the $SO(3)$ rotational symmetry is broken down to $SO(2)$ or completely. This may result from a constant external magnetic field as in the lattice QCD studies above, from unequal values of spin interactions in different directions as in certain spin models, or from the asymmetric initial conditions in the off-central heavy-ion collisions. 
In the latter case --- which is focus of this paper --- the anisotropy arises not from the underlying microscopic theory, QCD,  but from the different pressure gradients in different directions on the interaction plane. This translates into anisotropy in the VeV of the energy-momentum tensor in the plasma state while keeping its source, i.e. the metric, isotropic. It is tempting to ask whether the afore discussed inverse catalysis of the quark condensate can arise in this situation.
In this paper we will answer this question {\em in the affirmative} using the techniques of the gauge-gravity duality that are suitable for studying the effects of anisotropy in the full non-perturbative system.

The gauge/gravity duality is established as a powerful theoretical tool to study especially the qualitative aspects of a large class of strongly-coupled gauge theories in a completely non-perturbative manner. Several works have already approached the problem of the dependence of the condensate on the magnetic field in the holographic context, including \cite{Preis:2010cq,Preis:2012fh,Jokela:2013qya,Mamo:2015dea,McInnes:2015kec,
Evans:2016jzo,Gursoy:2016ofp,Nijs:2016pmh,Ballon-Bayona:2017dvv,Gursoy:2017wzz,Rodrigues:2017iqi,Rodrigues:2018pep,Rodrigues:2018chh}.
In \cite{Gursoy:2016ofp}, in particular, the authors gave a heuristic explanation of the IMC inspired by the aforementioned competition between the valence v.s the sea quarks but translated in the gravity language. Just as in (\ref{qqint}), there are two contributions that can be separately recognized \cite{Gursoy:2016ofp}  in the gravitational description as well. The first one, ``valence'' comes from explicit dependence of the open string tachyon equation of motion on $B$, that is the bulk field dual to the condensate, while the second, ``sea'' refers to an indirect effect coming from the backreaction of $B$ on the geometry. The authors of \cite{Gursoy:2016ofp} pointed out that is natural to identify the former explicit dependence with the valence, and the latter, implicit dependence with the sea contributions, respectively. If true, then, it would imply that the backreaction contribution is responsible for the IMC.

We consider a holographic theory dual to QCD with flat metric and no external magnetic field but with anisotropy. One way to introduce this is to turn on a relevant (or marginal) operator that (i) depends explicitly on one of the spatial directions and (ii) couples only to the color degrees of freedom. Indeed, this kind of deformation has been previously considered in the context of holography, e.g. in \cite{Mateos:2011ix,Mateos:2011tv,Giataganas:2017koz} for massless quark flavors. In these papers the authors considered a $\theta$-parameter (which sources the pseudo-scalar operator $\text{Tr}\,F\wedge F$) that depends linearly in one of the spatial directions, $\theta(x)=a\, x_3$, as a way to introduce anisotropy into the system.

Now let us see that the field theory with this spatially dependent $\theta$ term and massless quarks can also be put in a form similar to (\ref{DB}), hence the expectation value of the quark condensate can again be split into the valence and the sea parts as above. Consider the generating function of QCD both with a nontrivial $\theta$ term and an external axial gauge field $A_5$:
\be\label{gf2}
Z[A_5,\theta] = \int {\cal D}q\, {\cal D}A^a e^{-\int L[A^a,q] + A_5 \cdot J^5 + \theta \text{Tr} \star F \wedge F}\,
\ee
where $L[A^a,q]$ is the Lagrangian for the massless QCD and $J^5$ is the anomalous chiral current. We do not turn on an external electric gauge field for simplicity. Let us call the anomaly coefficient $c_a$ i.e. we have the non-conservation equation
\be\label{anomaly1}
d\star J_5 = c_a \, \text{Tr} F\wedge F\, .
\ee
This generating function enjoys invariance under the generalized chiral transformation\footnote{It is easy to realize this symmetry in the holographic dual by introducing a St\"uckelberg scalar coupled to the gauge field that corresponds to  $J_5$. We will not do this in this paper for simplicity.}
\be\label{sym1}
A_5\to A_5 + d \l_5, \qquad \theta \to \theta-c_a\l_5\, .
\ee
Therefore a nontrivial space dependent $\theta$ term, $\theta = a x_3$ can be removed by turning on an external axial gauge field $A_{5,\m}= a/c_a \delta_\m^3$. This means that the expectation value of the quark condensate in the theory with $\theta = a x_3$ and $A_5=0$, which we considered in the previous paragraph, can be rewritten as
\be\label{qqint3}
\langle\bar{q}q\rangle_{a}=\frac{1}{\mathcal{Z}(a)}\int \mathcal{D}A_\mu^a e^{-S_g}\det (\slashed{D}(a))\text{Tr}\,(\slashed{D}(a))^{-1}\,,
\ee
where
\be\label{Da}
\slashed{D}(a) = \gamma^{\mu}\left( \partial_\mu + A^a_\mu T^a\right) + \frac{a}{c_a} \gamma^3 \gamma^5  \,.
\ee
This is in the form (\ref{DB}) where the anisotropy enters the inverse propagator linearly and the contribution of $a$ to the quark condensate can be divided into the valence and the sea parts as above.

Anisotropic confining gauge theories were revisited in the holographic approach recently in \cite{Giataganas:2017koz}. One of the important lessons of this paper  was that, in the color sector, the anisotropic deformation reduces the confinement-deconfinement phase transition temperature. Since the effects of $B$ on confinement are qualitatively the same, this result reinforced the intuition of \cite{Gursoy:2016ofp}, and led to the conjecture that anisotropy by itself could explain the phenomenon of IMC.\footnote{A similar effect due to angular momentum, and dubbed as ``inverse shear catalysis'', was found in \cite{McInnes:2015kec}; we point out that angular momentum also induces anisotropy, which we will argue is the underlying physical reason behind all these phenomena.}

In order to  study the behavior of the chiral condensate $\langle\bar{q} q\rangle$ in the presence of anisotropy we have to consider an extra flavor sector on top of the model in \cite{Giataganas:2017koz}. Alternatively, we can introduce the same anisotropic deformation in the models originally considered in \cite{Gursoy:2016ofp}, at zero magnetic field. The difference between these two approaches boils down to the choice of potentials for the dilaton field. We choose to do the latter, because the choice of potentials is better motivated than in the former models. In this case, the color sector of the theory is taken to be Improved Holographic QCD (IHQCD) \cite{Gursoy:2007cb,Gursoy:2007er}. This is a bottom-up Einstein-Dilaton theory with a specific potential for the dilaton, which mimics many of the phenomenological signatures of QCD\@. On top of this theory, we also consider a flavor sector based on a pair of space filling $D4-\overline{D4}$ branes \cite{Bigazzi:2005md,Casero:2007ae}\@. However, since flavor physics is suppressed in the large $N_c$ limit, one must consider an appropriate limit in order to properly take into account the backreaction of flavors. Specifically, one must take both $N_c\to\infty$ and $N_f\to\infty$, while keeping their ratio $x=N_f/N_c$ fixed. This is known as the Veneziano limit, and defines the V-QCD model \cite{Jarvinen:2011qe} which is the model we use as the holographic dual of QCD in this paper.

The paper is organized as follows. In section \ref{sec:setup} we start by giving a brief overview of the model, discussing in detail the color and flavor sectors mentioned above, as well as presenting the relevant equations of motion and constraints. In section \ref{sec:IR} we discuss the IR asymptotics in detail and show, in particular, the drastic effects induced by the anisotropic deformation. In section \ref{sec:thermodynamics} we solve numerically the equations of motion and find the relevant anisotropic black brane solutions. We also study the thermodynamics of the models by working out the free energy in the canonical ensemble and discussing in detail the role of the anisotropic deformation. In section
\ref{sec:obs} we compute various observables of physical interest. First, we devote our attention to study the chiral condensate, which was the original motivation of the paper. In addition, we study the meson and glueball spectra, quark-antiquark potential and entanglement entropy, to further characterize the behavior of the new IR fixed points.
We close in \ref{sec:con} with a discussion of our results and some outlook.

\section{Holographic setup}\label{sec:setup}

The holographic model we will consider has two parts, the gluon sector and the flavor sector. The gluon sector is based on the so-called improved holographic QCD model (IHQCD) \cite{Gursoy:2007cb,Gursoy:2007er}, and the flavor sector is defined in terms of a generalized tachyon Dirac-Born-Infeld action arising from a pair of space filling $D4-\overline{D4}$ branes~\cite{Bigazzi:2005md,Casero:2007ae}. The two actions fully backreact in the Veneziano limit, which defines the V-QCD model~\cite{Jarvinen:2011qe}:
\be
S=S_g+S_f\,,
\label{action}
\ee
where
\be
S_g= M^3 N_c^2 \int d^5x \ \sqrt{-g}\left(R-{4\over3}{
(\partial\lambda)^2\over\lambda^2}+V_g(\lambda)-\frac{1}{2}Z(\lambda)(\partial \chi)^2\right) \,,
\label{actg}
\ee
and
\begin{align}
\label{actf}
S_f =-x\, M^3 N_c^2 \int d^5x V_f(\l,\t) \sqrt{- \mathrm{det}\left(g_{\m\n} + \kappa(\l)\, \partial_{\m} \t \,\partial_{\n} \t\right) } \,.
\end{align}

The gluon sector contains a finite set of bulk fields dual to relevant or marginal operators that dominate the dynamics in the IR. Among these we have the stress-energy tensor $T_{\mu\nu}$, which is dual to the metric $g_{\mu\nu}$, the glueball operator $\text{Tr}\, F^2$ dual to the dilaton $\l$ and a pseudo-scalar operator $\text{Tr}\,F\wedge F$ dual to the axion $\chi$. The latter operator is introduced in order to break isotropy as in \cite{Mateos:2011ix,Mateos:2011tv,Giataganas:2017koz}.  Notice that proper implementation of the QCD axial anomaly would require coupling of the axion to the flavor sector which is of the leading order in the Veneziano limit~\cite{Arean:2013tja,Arean:2016hcs}. As we will only use the axion to break the isotropy, considering such couplings is not necessary and we omit them for simplicity. Finally, the flavor sector includes an additional field, the tachyon $\t$, which is dual to the quark bilinear operator $\bar{q} q$.

Constraints to the potential functions and couplings in the action from various sources have been discussed in detail in earlier literature~\cite{Gursoy:2007cb,Gursoy:2007er,Gursoy:2009jd,Gursoy:2012bt,Jarvinen:2011qe,Alho:2012mh,Arean:2013tja,Jarvinen:2015ofa,Arean:2016hcs,Jokela:2018ers}. In the current study, the  coupling $Z(\lambda)$ between the dilaton and the axion is taken from \cite{Gursoy:2012bt,Drwenski:2015sha} while the other potentials
are taken from \cite{Alho:2012mh,Alho:2013hsa}. Explicitly, the potentials are given by
\begin{eqnarray}
\label{Vf0SB}
V_g(\lambda)&=&{12\over \mathcal{L}_0^2}\biggl[1+{88\lambda\over27}+{4619\lambda^2
\over 729}{\sqrt{1+\ln(1+\lambda)}\over(1+\lambda)^{2/3}}\biggr]\, , \\
 V_{f}(\lambda,\tau)& =& {12\over x \mathcal{L}_{UV}^2}\biggl[{\mathcal{L}_{UV}^2\over\mathcal{L}_0^2}
-1+{8\over27}\biggl(11{\mathcal{L}_{UV}^2\over\mathcal{L}_0^2}-11+2x \biggr)\lambda\nn\\
 &&+{1\over729}\biggl(4619{\mathcal{L}_{UV}^2\over \mathcal{L}_0^2}-4619+1714x - 92x^2\biggr)\lambda^2\biggr]\,e^{-a_0\tau^2} \, , \\
 \kappa(\l) &=& {[1+\ln(1+\l)]^{-1/2}\over[1+\frac{3}{4}(\frac{115-16x }{27}-{1\over 2})\l]^{4/3}}\, ,\label{kappaa}\\
Z(\lambda)&=&1+\frac{\lambda ^4}{10}\,, \label{Zdef}
 \end{eqnarray}
where
\be
 a_0 = \frac{3}{2\mathcal{L}_{UV}^2}\, ,\qquad\mathcal{L}_{UV}^3 = \mathcal{L}_0^3 \left( 1+{7 x \over 4} \right) \, .
\label{adsrad}
\ee
The parameter $\mathcal{L}_0$ is the AdS radius for $x=0$, which we set to one in our numerics. In a similar fashion, $\mathcal{L}_{UV}$ is the AdS radius for $x\neq0$. Notice that we have set the overall constant in $Z(\lambda)$ ($Z_0$ in the notation of \cite{Gursoy:2012bt}) to unity, since it can be reabsorbed in the normalization of $\chi$.

Our Ansatz for the metric and other bulk fields is the following:
\begin{equation}\label{ansatz}
\begin{split}
&ds^2=e^{2A(r)}\left[-f(r) dt^2+dx_1^2+dx_2^2  + e^{2W(r)} dx_3^2+\frac{dr^2}{f(r)}\right]\,,\\
&\qquad\quad\,\,\lambda=\lambda(r)\,,\qquad\chi=a\, x_3\,,\qquad\tau=\tau(r)\,.
\end{split}
\end{equation}
This Ansatz automatically satisfy the equations of motion for the axion $\chi$, while introducing anisotropy in the $x_3$ direction. Moreover the dependence of metric field $W$  on the holographic coordinate $r$ precisely corresponds to the dependence of the renormalized anisotropy parameter on the RG scale discussed in the introduction.

The Einstein equations that follow from the action are:
\begin{align}
&3 A'' +{2\l'^2 \over 3\l^2} + 3 A'^2 + \left(3 A' -W'\right){ f' \over 2f}    -{ e^{2 A} V_g(\l) \over 2 f} +{x G e^{2 A}V_f(\l,\t) \over  2 f}-\frac{a^2 e^{-2 W} Z(\lambda)}{4 f}=0 \, , \nn \\
& \qquad\qquad\qquad\quad\,\, W''+ {W' f'\over f} +(3 A'+W')W'  +\frac{a^2 e^{-2 W} Z(\lambda)}{2 f}=0 \, , \label{backeq} \\
&\qquad\qquad\qquad\qquad\qquad\qquad\,\,\,\, f''+(3 A'+  W') f'  =0 \,,\nn
\end{align}
where we have defined
\begin{align}
 \label{Gdef}
 G(r) \equiv \sqrt{1 + e^{-2 A(r)}\kappa(\l) f(r) \t'(r)^2}\,.
\end{align}
There is also a first order constraint, which is given by
\begin{align}
&{2\l'^2 \over 3\l^2}- \left(3  A' +W' \right) {f' \over 2  f}-3A'\(2A' +W'\) +{ e^{2 A} V_g(\l)\over 2 f}  -{x  e^{2 A}V_f(\l,\t) \over 2  G  f} -\frac{a^2 e^{-2 W} Z(\lambda)}{4 f}=0 \,.
\end{align}
The equation of motion for the dilaton is
\begin{align}\label{laeq}
& {\l'' \over \l} -{\l'^2 \over \l^2}+\left(  3 A' +W' +   {f' \over f}   \right) {\l' \over \l} + {3\l  e^{2 A} \over 8f}  \partial_{\l} V_g(\l)-{3  x  e^{2  A}   G  \l \over 8 f } \partial_{\l} V_f(\l,\t)\nn \\
&\qquad\qquad\quad-{3  x  \l  V_f(\l,\t)  \t'^2 \over 16  G } \partial_{\l} \kappa(\l)-\frac{3 a^2 \lambda e^{-2 W}}{16 f}\partial_{\l} Z(\l)=0 \,.
\end{align}
Finally, the equation of motion for the tachyon is
\begin{align} \label{tacheq}
&\t''-{e^{2 A}  G^2 \over f \kappa(\l)}{\partial_{\t} \log  V_f (\l,\t)}+ e^{-2  A}  f  \k(\l)  \bigg(4 A'+W' +{ f' \over 2f}  +{\l' \over 2}  \partial_{\l} \log ( V_f(\l,\t)^2 \kappa(\l) ) \bigg) \t'^3 \nn \\
&\qquad\qquad\qquad\quad\,\,\,\,+ \left(3A'+W' +{f' \over f} + \l' \partial_{\l} \log(V_f(\l,\t) \k(\l)) \right) \t' =0 \,.
\end{align}
We will solve this set of equations numerically.
To do this, we make use of a scale symmetry present in the equations of motion:
\be \label{scalesymmetry}
r \mapsto r\tilde\Lambda, \qquad A \mapsto A - \log\tilde\Lambda, \qquad  a \mapsto \frac{a}{\tilde\Lambda}.
\ee
This corresponds to the scale symmetry of the action on the field theory side.
The background solution has a nontrivial dependence on the bulk coordinate $r$, which breaks this symmetry and introduces an energy scale $\Lambda$, analogous to $\Lambda_\text{QCD}$\@.
The precise definition of $\Lambda$ will be given in section \ref{sec:thermodynamics}\@.
Before discussing the numerical solutions, we will study in detail the IR structure of these equations and obtain analytic solutions in this regime.

\section{IR behavior \label{sec:IR}}

Let us then discuss the asymptotic geometry and RG flow at zero temperature (i.e., for $f=1$) in the IR. As it turns out, turning on any nonzero anisotropic parameter $a$ changes the IR structure drastically. Motivated by the fact that in the isotropic case the quarks are either asymptotically decoupled in the IR in the chirally broken phase or affect the gluon dynamics only trivially in the symmetric phase~\cite{Jarvinen:2011qe,Arean:2013tja}, we start by considering the system without quarks, i.e., taking the limit $x \to 0$ above.

It is useful to write the equations of motion in another form. We define $\l =e^\phi$, $\frac{dr}{dA} e^A =q = -e^p$, and $\widetilde W =W +A$. In terms of them, the three independent equations of motion are
\begin{align}
 8 \dot \phi^2 &=e^{2 p} \left(3 a^2 e^{-2 \widetilde W} Z(\phi )-6 V_g(\phi )\right)+36 \left(\dot{ \widetilde W}+1\right) &\\
 \dot p&=\frac{1}{6} \left(-2 e^{2 p} V_g(\phi )+6 \left(\dot{\widetilde W}-1\right)+24\right)& \label{peq} \\
 \ddot {\widetilde W}&=-\frac{1}{6} e^{2 p} \left(3a^2e^{-2 \widetilde W} Z(\phi )+2  V_g(\phi ) \left(\dot{\widetilde W}-1\right)\right) \ .&
\label{Wteq}
 \end{align}
where dots denote the derivatives with respect to $A$.

\subsection{AdS$_4$ IR fixed point in the chirally symmetric phase \label{sec:FP}}

We start by discussing the exact fixed point solutions which are realized in the chirally symmetric  phase in the zero temperature limit.
First we notice that the above equations of motion seem to admit an exact fixed point solution, determined by the equations
\be
e^{2p_*} V_g(\lambda_*)=9 \ , \qquad 3 a^2 Z(\lambda_*)=2 V_g(\lambda_*) \label{FPsol}
\ee
and with $\widetilde W$ a constant which we set to zero (as it can be absorbed in $a$). This fixed point (without additional requirements) is however not realized as an endpoint of any holographic RG flow. The reason can be seen as follows. The second order dilaton EoM may be written as
\begin{align}
 12 \ddot \phi \dot \phi &=\frac{9}{4} e^{2 p} \dot \phi \left[a^2 e^{-2 \widetilde W} Z'(\phi )-2 V_g'(\phi )\right]& \nn\\
  & \ \ \ + e^{2 p} V_g(\phi ) \left[e^{2 p}\left(-\frac{3}{2}a^2e^{-2 \widetilde W} Z(\phi )+  V_g(\phi )\right)+\left(2 e^{2 p} V_g(\phi ) -18\right)-18 \dot{\widetilde W}\right] \ . &
\end{align}
The term in the latter square brackets vanishes at the fixed point (because also $\dot{\widetilde W} =0$) but the term in the first square brackets does not. After dividing by $\dot \phi$, the first two terms already imply that $\ddot\phi$ is finite. So even if the EoM is satisfied when $\dot\phi$ vanishes exactly, any small perturbation will lead to a sizable $\ddot\phi$ and therefore fast deviation from the fixed point. In particular, the fixed point cannot be reached asymptotically for $A \to \pm \infty$.

If in addition to the conditions~\eqref{FPsol} we impose (again taking $\widetilde W = 0$)
\be
 a^2 Z'(\phi_* ) = 2 V_g'(\phi_* )
\ee
the  issue with the flow of the dilaton is gone and the fixed point is stable and physical. Notice that this condition can also be written as
\be \label{extracond}
 \frac{d}{d\phi}\log Z(\phi_* ) =3 \frac{d}{d\phi}\log V_g(\phi_* )
\ee
so that the flow equations~\eqref{phiflow}--\eqref{aeqflow} also have trivial solutions. The additional condition cannot however be satisfied for a generic $a$, but only for some specific value which we denote by $a_*$. As one can check, for the potentials~\eqref{Vf0SB}--\eqref{kappaa} there is no such solution (excluding the runaway solution at $\phi_* = \infty$).

The situation is different if we consider the backreaction of the flavors in the chirally symmetric phase, $\tau=0$. In this case the EoMs for the glue are obtained by replacing $V_g$ by the effective potential $V_\mathrm{eff} = V_g - x V_{f0}$~\cite{Jarvinen:2011qe}. As it turns out, after the replacement and for the potentials specified above, a nontrivial fixed point solution $(\phi_*,p_*,a_*)$ does exist for $x \lesssim 1$. As we show below, this fixed point is indeed realized the IR limit of the $T=0$ RG flows in the symmetric phase for $x=1/3$.

As $p_*$ is fixed, the resulting geometry is AdS$_4 \times \mathbb{R}$: $e^{A(r)}/A'(r) = - e^{p_*}$ is solved by $e^A = e^{p_*}/r$, and the warp factor of $dx_3^2$ is $e^{\widetilde W} = e^{A+W} = \mathrm{const.}$ as we pointed out above.

\subsection{Rolling IR fixed point in the tachyonic phase\label{sec:rolling}}

The low temperature geometries in the chirally broken phase (and also for the pure glue case $x=0$) have an interesting structure which is drastically different from that of the isotropic solutions.
In order to analyze them, we start by studying variations around the exact fixed point discussed above, which leads to a ``slow roll'' behavior as we will now demonstrate.
Without loss of generality, we may take $\widetilde W(A=0) =0$ and discuss the evolution of the system near $A=0$. The ``slow roll'' will be driven by Eq.~\eqref{peq}\@. We substitute the following Ansatz in the equations of motion:
\be
 p = \hat p_*+C_p A\ ,\qquad \widetilde W = C_W A\ , \qquad \phi = \hat \phi_* + C_\phi A \ ,
\ee
where all coefficients are taken to be small and the fixed point values may depend on them, $\hat p_*=\hat p_*(C_i)$ and $\hat \phi_*=\hat \phi_*(C_i)$. As it turns out, the EoMs are satisfied to first order in the coefficients $C_i$ if the proportionality $Z(\phi) \,\propto\, V_g(\phi)^3$ holds at least for the exponential terms in $\phi$, in consistency with~\eqref{extracond}. The slowly rolling functions $p(A)$ and $\phi(A)$ satisfy the equations
\be
 e^{2p(A)} V_g(\phi(A))=9\left(1+\frac{C_W}{2}\right) \ , \qquad a^2 e^{-2C_W A } Z(\phi(A)) = \frac{2}{3}\left(1-C_W\right) V_g(\phi(A))
\ee
up to corrections $\mathcal{O}(C_i^2)$. This implies in particular that $C_p = - C_W/2$ and that $C_\phi = C_W V_g(\phi_*)/V_g'(\phi_*)$. The conditions for $\hat p_*$ and $\hat \phi_*$ are obtained by setting $A=0$ and they are corrected by $\mathcal{O}(C_W)$ terms with respect to~\eqref{FPsol}. Therefore $\hat p_*$ and $\hat \phi_*$ approach the fixed point values $p_*$ and $\phi_*$ defined by~\eqref{FPsol} as $C_W \to 0$. The value of $C_W$ can be related to the deviation from the law $Z(\phi) \,\propto\, V_g(\phi)^3$ but this requires considering higher order corrections in $C_W$ which we will not do here. Instead, we will present a more precise way of discussing the IR flow as follows.

The IR flow will actually determined by an fixed point which is slowly moving due to the flow of, say $\widetilde W$. In order to guarantee that the flow stays at the fixed point, it is sufficient, to a very good precision, to simply set all second derivatives with respect to $A$ to zero.
Notice that~\eqref{Wteq} then can be solved for $\widetilde W'$, and the system can be rearranged to read
\begin{align}
 8 \phi '^2 &=3 a^2 e^{2 p-2 \widetilde W} Z(\phi )-\frac{54 a^2 e^{-2 \widetilde W} Z(\phi )}{V_g(\phi )}-6 e^{2 p} V_g(\phi )+72 &\\
 p'&=-\frac{3 a^2 e^{-2 \widetilde W} Z(\phi )}{2 V_g(\phi )}-\frac{1}{3} e^{2 p} V_g(\phi )+4& \label{peqn} \\
 \widetilde W'&=1-\frac{3 a^2 e^{-2 \widetilde W} Z(\phi )}{2 V_g(\phi )} \ .&
\label{Wteqn}
\end{align}
For these to lead to a consistent flow (rather than being satisfied only at a single point), we require that their derivatives are also satisfied, imposing the approximation that second derivative with respect to $A$ are set to zero. This leads to two new equations which are equivalent to the above equations at certain values of $\phi$ and $q$. This then determines the flow. Since the potentials are possibly complicated functions of $\phi$, it is convenient to solve $a$ (or actually $a e^{-\widetilde W}$) and treat $\phi$ as a parameter. 
That is, the above three equations and their derivatives lead to five independent equations which we solve for $\phi'$, $p'$, $\widetilde W'$, $p$, and $ae^{-\widetilde W}$. There is a trivial solution without flow given by Eqs.~\eqref{FPsol}. The nontrivial solution is given by
\begin{align}
\label{phiflow}
 \phi '&= 
 \frac{6 \left(\frac{d}{d\phi}\log Z(\phi )-3 \frac{d}{d\phi}\log V_g(\phi )\right)}{\mathcal{D}} & \\
\label{Wflow}
 \widetilde W'&= 
 \frac{3 \left(\frac{d}{d\phi}\log V_g(\phi )-\frac{d}{d\phi}\log Z(\phi )\right) \left(3 \frac{d}{d\phi}\log V_g(\phi )-\frac{d}{d\phi}\log Z(\phi )\right)}{\mathcal{D}} &\\
\label{pflow}
 p'&=  
 \frac{3 \frac{d}{d\phi}\log V_g(\phi ) \left(3 \frac{d}{d\phi}\log V_g(\phi )-\frac{d}{d\phi}\log Z(\phi )\right)}{\mathcal{D}} &\\
\label{peqflow}
 e^{2p}&= 
 \frac{18  \left(8-3 \frac{d}{d\phi}\log V_g(\phi ) \frac{d}{d\phi}\log Z(\phi )+2 \left(\frac{d}{d\phi}\log Z(\phi )\right)^2\right)}{ V_g(\phi)\mathcal{D}} &\\
\label{aeqflow}
a^2e^{-2\widetilde W}&= 
 \frac{2 V_g(\phi) \left(16+9 \frac{d}{d\phi}\log V_g(\phi ) \frac{d}{d\phi}\log Z(\phi )-9 \left(\frac{d}{d\phi}\log V_g(\phi )\right)^2\right)}{3 Z(\phi) \mathcal{D}} \ ,&
\end{align}
where
\be
 \mathcal{D} = 16 -3 \frac{d}{d\phi}\log V_g(\phi ) \frac{d}{d\phi}\log Z(\phi ) +3 \left(\frac{d}{d\phi}\log Z(\phi ) \right)^2 \ .
\ee
Remarks:
\begin{itemize}
 \item The equations~\eqref{phiflow}--\eqref{pflow} are consistent with the relations of the coefficients $C_i$ obtained above in Sec.~\ref{sec:FP} when  $Z(\phi)$ is roughly $\propto\, V_g(\phi)^3$, i.e., $\frac{d}{d\phi}\log Z(\phi )-3 \frac{d}{d\phi}\log V_g(\phi )$ is small. Now the value of $C_W$ is fixed and can be read from~\eqref{Wflow}.
 \item The other equations~\eqref{peqflow} and~\eqref{aeqflow} generalize the fixed point equations~\eqref{FPsol}.
 \item For the flow to present an acceptable IR asymptotics we need such potentials that $\phi'<0$ in~\eqref{phiflow}.
\item Interestingly, inserting to the flow equations the exactly exponential potentials $V=V_0 e^{\sqrt{8/3} \sigma \phi}$, $Z =e^{2\sqrt{8/3}\gamma\phi}$ reproduces the exact scaling solution of~\cite{Giataganas:2017koz} even when deviation from $Z(\phi)\, \propto\, V_g(\phi)^3$ is sizable. This is because corrections to the flow involve higher order logarithmic derivatives which vanish for exactly exponential potentials.

\end{itemize}

\begin{figure}[t!]
\centering
\includegraphics[width=0.5\textwidth]{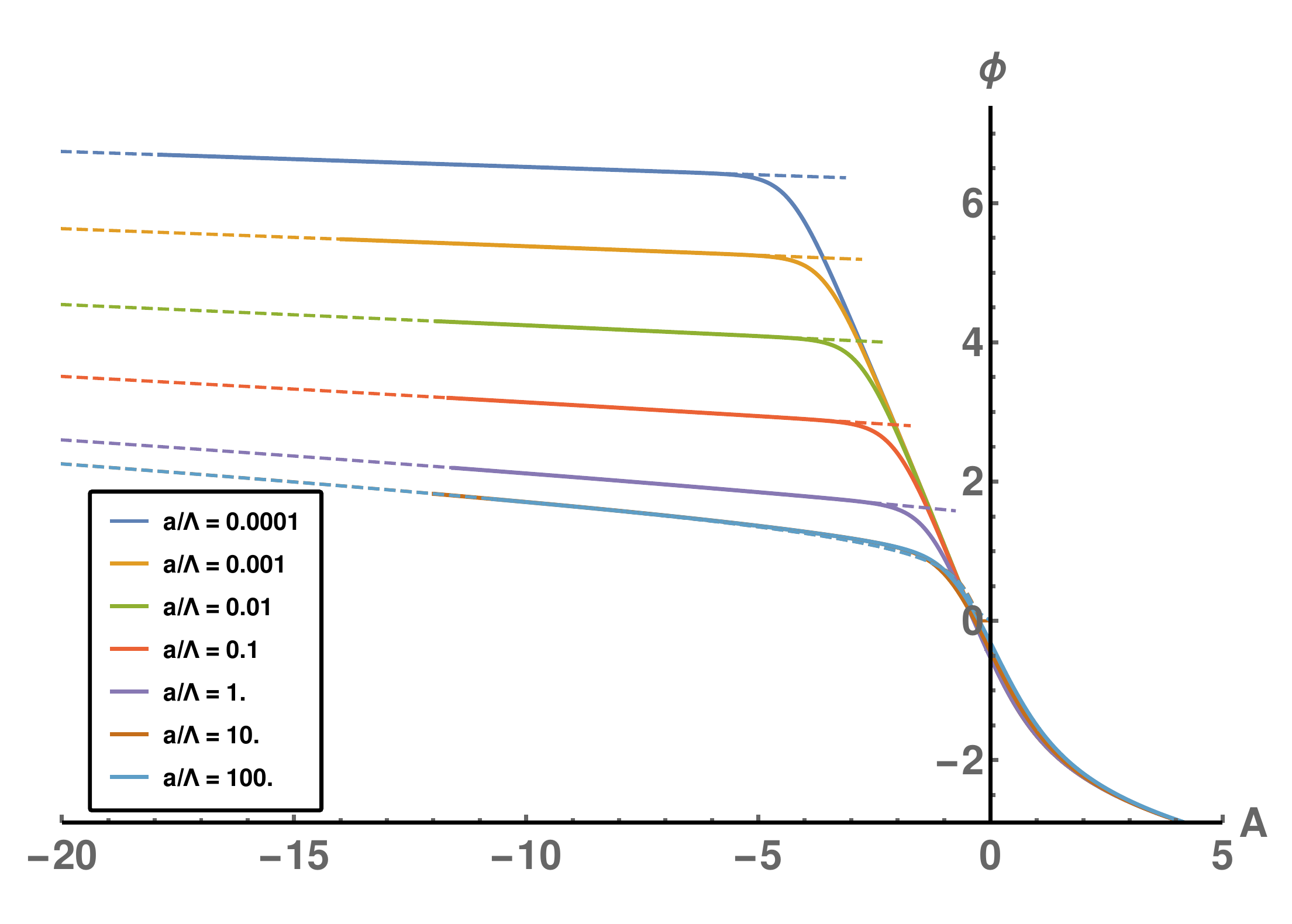}%
\includegraphics[width=0.5\textwidth]{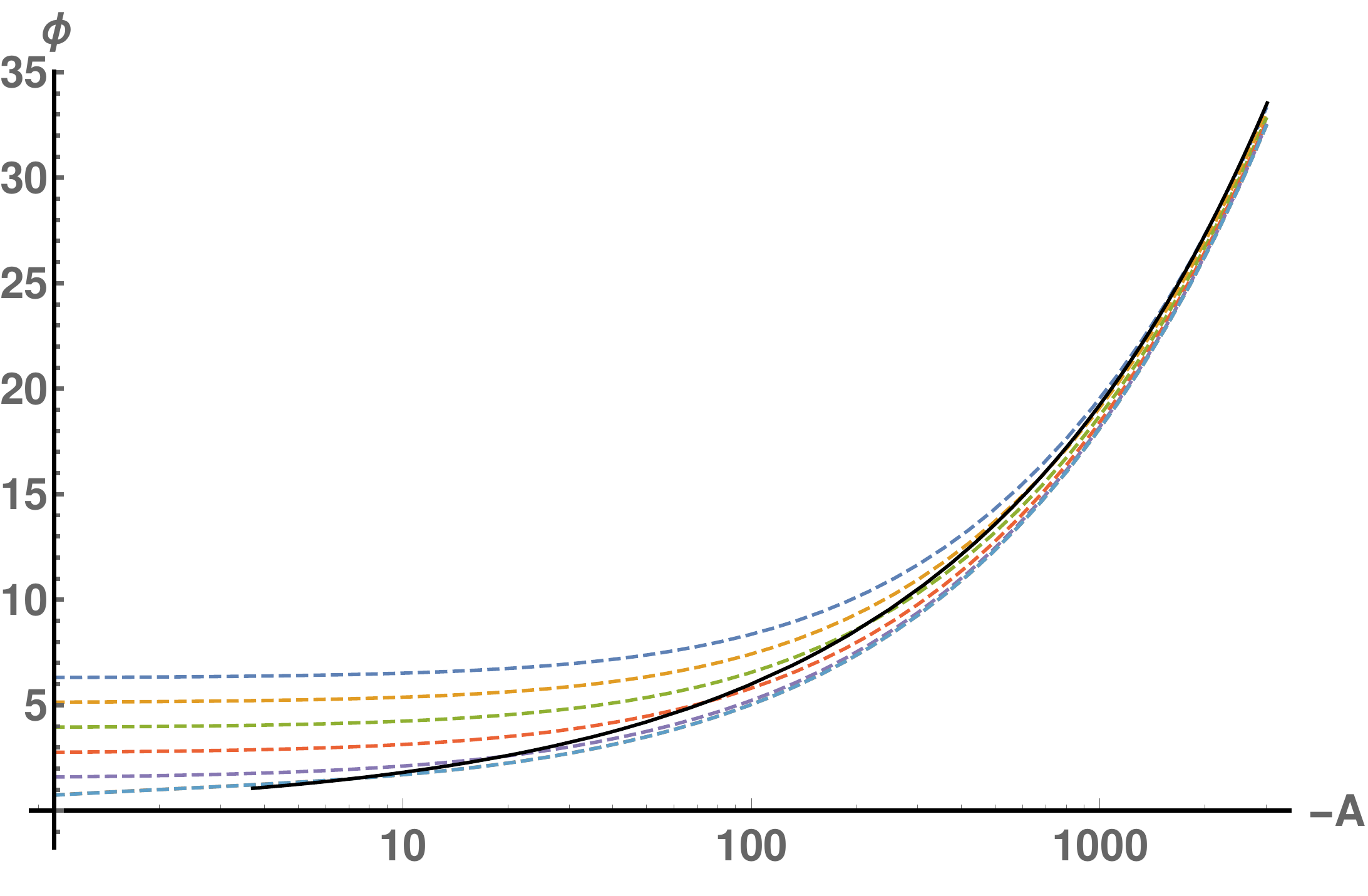}
\caption{\label{fig:IRRGflow}The holographic RG flow of the coupling in the IR regime at $x=0$. Left: The exact numerically constructed flow (solid curves) compared to the flow given by Eq.~\protect\eqref{phiflow}  (dashed curves) for several values of $a$. Right: The RG flows for various $a$ (dashed curves, computed from~\protect\eqref{phiflow}) compared to the asymptotic result of Eq.~\protect\eqref{IRasympt} (black solid curve).}
\end{figure}

\subsubsection{Behavior at asymptotically large $\l$}

The potentials in Eqs.~\eqref{Vf0SB}--\eqref{Zdef} are such that the IR solutions are very precisely described by the above flow equations because the proportionality $Z(\phi)\, \propto\, V_g(\phi)^3$ is only violated by logarithmic corrections.

We first discuss the asymptotic IR behavior of the geometry which can be solved analytically by using the flow equations.
This has been worked out for more generic potentials and higher order corrections in \cite{AriasTamargo:2018pcx}\@.
In order to find the asymptotics, we substitute  the potentials in the flow equations,
i.e., we take $V_g \,\propto\, V_\mathrm{IR} e^{4\phi/3}\sqrt{\phi}$ and $Z \,\propto\, Z_\mathrm{IR} e^{4\phi}$. We obtain
\begin{align}
 \phi '(A) &\simeq -\frac{3}{16 \phi(A)}\ ,\quad
\, \quad  e^{2 p(A)} &\simeq \frac{9 e^{-\frac{4 \phi(A)}{3}}}{V_\mathrm{IR} \sqrt{\phi(A)}}\ ,\quad a^2e^{-2\widetilde W(A)}\simeq \frac{2 V_\mathrm{IR} e^{-\frac{8 \phi(A)}{3}}  \sqrt{\phi(A)}}{3 Z_\mathrm{IR} }&
\end{align}
up to corrections suppressed by $1/\phi$.
Integrating these equations, we obtain the asymptotics for the metric factors and $\phi$ as $r \to \infty$:
\be \label{IRasympt}
 e^{A} \sim \frac{1}{r}e^{-\sqrt{(\log r)/6}-(\log\log r)/8} \ , \quad e^{W+A} = e^{\widetilde W} \sim e^{\sqrt{(2\log r)/3}-(\log\log r)/8} \ , \quad \phi \sim \sqrt{(3\log r)/8} \ .
\ee

These formulas describe an approximate AdS$_4 \times \mathbb{R}$ geometry with multiplicative corrections of the form $e^{\#\sqrt{\log r}}$. It is instructive to write down the asymptotic string frame metric as $r \to \infty$:
\be \label{SFmetric}
ds_s^2 = e^{4\phi/3}ds_E^2 \sim \frac{1}{r^2}(\log r)^{-1/4}\left[ - dt^2+dx_1^2+dx_2^2 + dr^2\right] + e^{\sqrt{6\log r}}(\log r)^{-1/4}dx_3^2 \ .
\ee
Notice the cancellation of the square roots in the warp factor, after which the first term is the AdS$_4$ metric with multiplicative logarithmic corrections.

\subsubsection{Numerical analysis of the IR RG flows}\label{sec:numericalIR}

Numerically solving the flow equations~\eqref{phiflow}--\eqref{aeqflow} leads to an accurate description of the IR behavior of the model. We demonstrate this in Fig.~\ref{fig:IRRGflow} (left), where we compare numerically the RG flow of the coupling obtained by solving the EoSs exactly to that given by the flow equations for the pure glue case ($x=0$) and various values of the asymmetry parameter $a$. Recall that the IR (UV) is at $A \to -\infty$ ($A \to +\infty$). The exact  numerical solutions of the full equations of motion are given by the solid curves whereas the flows from Eq.~\eqref{phiflow} are given as the dashed curves. We took the initial conditions for the latter flows from the exact solutions at $A=-10$.  When $a/\Lambda \ll 1$, the structure of the solution is as follows.
\begin{itemize}
 \item For $A\gg 0$, the geometry has the same UV asymptotics as in the absence of asymmetry~\cite{Gursoy:2007cb,Gursoy:2007er}, i.e., the RG flow is given by $\phi \sim -\log A$.
 \item For $A \ll 0$ and when $\phi \ll \phi_*$ there is an intermediate regime where the background follows the ``standard'' IR asymptotics $\phi \sim \frac{3}{2} A$ of the $a=0$ case~\cite{Gursoy:2007cb,Gursoy:2007er}. Here the fixed point value $\phi_*$ is given by Eqs.~\eqref{FPsol}: for small $a$ we have that $\phi_* \sim - 3(\log a)/4$.
 \item For $A \ll 0$ and $\phi \gtrsim \phi_*$ the flow is the ``slow roll'' described by Eq.~\eqref{phiflow}. As $A$ decreases, after a short transition regime, the exact numerical solution is practically indistinguishable from that given by Eq.~\eqref{phiflow}.
\end{itemize}
When $a=\mathcal{O}(1)$ the intermediate region is absent. As $a \to 0$, the region with the intermediate behavior grows. The ``slow roll'' region is pushed deeper and deeper in the IR and disappears in the limit. Therefore the geometry approaches the $a=0$ thermal gas geometry, but the convergence is not uniform.

\begin{figure}[t!]
\centering
\includegraphics[width=0.5\textwidth]{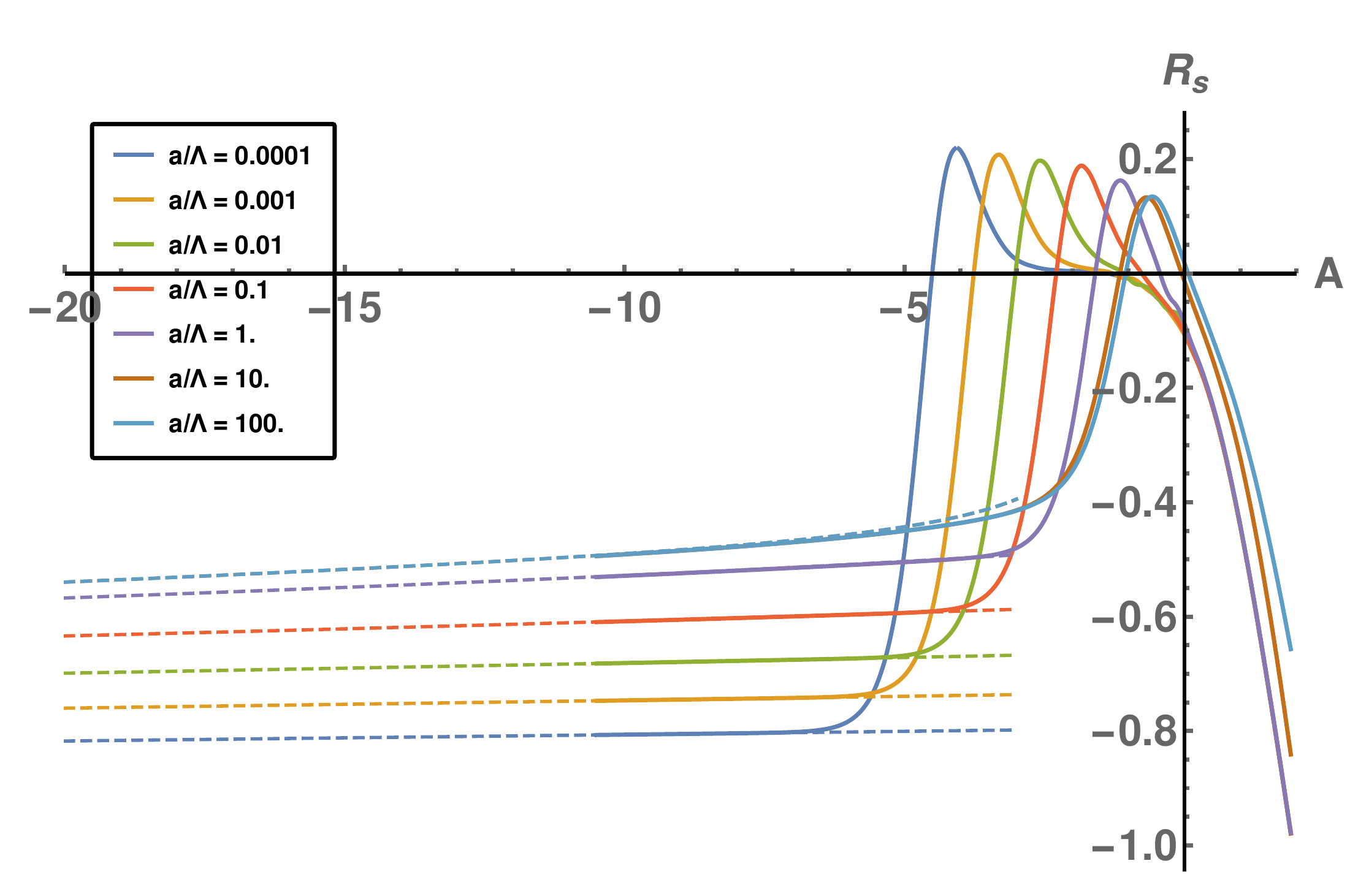}%
\includegraphics[width=0.5\textwidth]{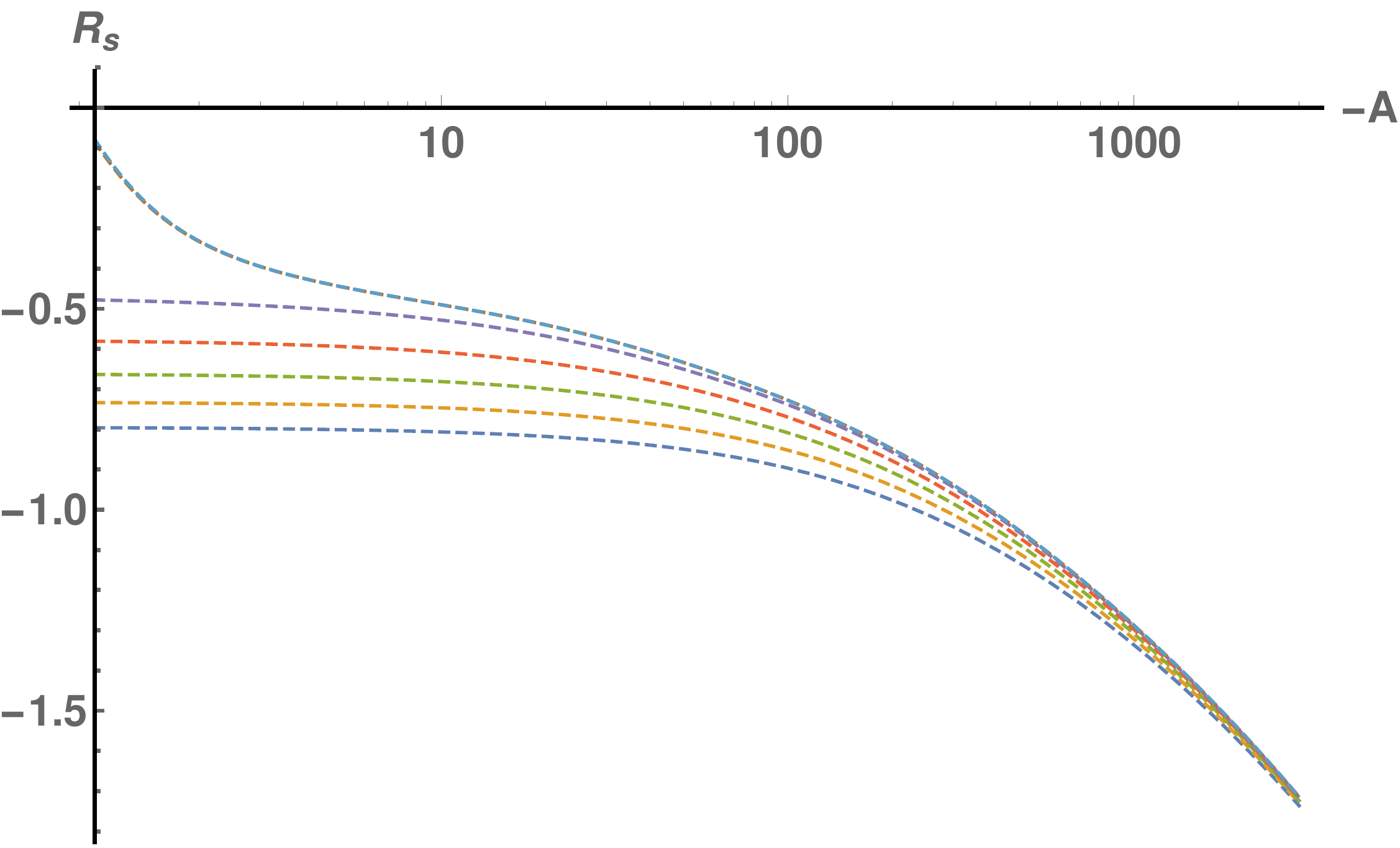}
\caption{\label{fig:Ricci} The string frame Ricci scalar as a function of $A$ for several values of $a$. Notation as in Fig.~\protect\ref{fig:IRRGflow}.}
\end{figure}

In Fig.~\ref{fig:IRRGflow} (right) we compare the RG flow of the coupling given by Eq.~\eqref{phiflow} and to the asymptotic formula in Eq.~\eqref{IRasympt} for various ``boundary conditions'' near $A = 0$, as determined by the value of $a$. The color coding for the values of $a$ is the same as in the left hand plot. We notice that the analytic asymptotic formula works reliably only at extremely high values of $-A$.

Next, we make the following consistency checks on the discussion above. In Appendix~\ref{app:tachyonIR} we verify that for the potentials used in this article the tachyon indeed diverges fast enough in the IR for it to decouple so that the above analysis is consistent even for the tachyonic solutions at $x>0$.

We also plot the string frame Ricci scalar which governs the higher order stringy corrections in Fig.~\ref{fig:Ricci}. We see that $|R_s|$ remains small in the IR in our region of interest so that the corrections are small and our approach is consistent. Notice however that at extremely deep in the IR the Ricci scalar grows as shown on the right hand plot and eventually blows up. This behavior is due to the logarithmic term in the AdS part of~\eqref{SFmetric}. Therefore the zero temperature geometries corresponding to the flow would receive large stringy corrections. Notice also that the limit of $a \to 0$ appears nonuniform as one can see from the left hand plot in Fig.~\ref{fig:Ricci}: as $a$ decreases the amplitude of $R_s$ in the IR grows but the ``slow roll'' regime, where $R_s$ is sizable, is pushed deeper in the IR.

Finally, it is important to understand what the divergence of the Ricci scalar means for the finite temperature solutions which we consider below. Since the IR geometry is close to AdS$_4$, small black holes will have temperatures $T \sim 1/r_h \sim e^{A_h}$ where the subscript ``h'' refers to the value at the horizon. Therefore the finite temperature geometries are consistent, that is free of stringy corrections, down to exponentially small temperatures: for example the Ricci scalar reaches the value $-2$ around $A \sim -4000$ which gives the temperature $T/\Lambda \sim 10^{-1700}$. We conclude that we can trust the finite temperature solutions down to an extremely small temperature scale, which for any practical application can be taken to be zero.

\section{Thermodynamics}\label{sec:thermodynamics}

To investigate thermodynamics of the system we consider the black brane solutions with a nontrivial blackening factor $f=f(r)$ in (\ref{ansatz}) which vanishes at some horizon $r_h$, $f(r_h)=0$. $r_h$ then parametrizes the different black-brane solutions with different temperatures $T = -f'(r_h)/4\pi$. We solve the equations of motion (\ref{backeq}), (\ref{laeq}) and (\ref{tacheq}) numerically by shooting from
the  horizon  towards  the  boundary. More specifically, we set the data of the non-normalizable modes according to
\be\label{boundaryassymp}
\begin{split}
A(r)\to-\log r \,,\qquad f(r)\to1\,,\qquad W(r)\to0\,,\qquad\\
\lambda(r)\to-\frac{1}{b_0\log(r\Lambda)}\,,\qquad\tau(r)\to m_q r(-\log(r\Lambda))^{-\frac{\gamma_0}{b_0}}\,,
\end{split}
\ee
with $b_0\equiv\frac{1}{3}(11-2x)$ and $\gamma_0\equiv\frac{3}{2}$. We shoot from the horizon until the desired value of
$m_q$ is reached -- the sources for the other fields can then be set to be those of~\eqref{boundaryassymp} by using symmetries of the solution~\cite{Jarvinen:2011qe,Gursoy:2016ofp}. In particular, the symmetry Eq.~\eqref{scalesymmetry} can be used to set\footnote{For the source $\Lambda$ to be precisely defined, we actually also need to specify the NLO terms (see for example~\cite{Jarvinen:2011qe})\@.} the value of the source $\Lambda$ of the dilaton field, which characterizes the energy scale of the solution (in analogy to $\Lambda_{QCD}$ on the field theory side). From  each  solution  we read off the value of the normalizable modes and extract from them the expectation values of the operators dual to the various bulk fields. For simplicity we will focus only on solutions with $m_q=0$. Other values of $m_q$ are straightforward to obtain but the extraction of the quark condensate is numerically more demanding in these cases.

In order to study the thermodynamics of our solutions we also need to compute the
free energy $F$ in the canonical ensemble. In the holographic description, this amounts to evaluating the on-shell Euclidean  action  (\ref{action}),
appended by the standard Gibbons-Hawking and counterterms \cite{Gursoy:2007er,Jarvinen:2015ofa}. In practice, however, we compute the background subtracted
free energy directly by integrating the first law of thermodynamics
\be
dF=-sdT\,,
\ee
while keeping all sources fixed. Before proceeding further it is worth noting that more generally, given the bulk fields of our gravitational action, the Ward identity for the boundary stress-energy tensor reads:
\be\label{Ward}
\partial^i\langle T_{ij}\rangle+\langle \mathcal{O}_\lambda\rangle\partial_j \lambda^{(0)}+\langle \mathcal{O}_\chi\rangle\partial_j \chi^{(0)}+\langle \mathcal{O}_\tau\rangle\partial_j \tau^{(0)}=0\, ,
\ee
where ${\cal O}_\a$ is the operator dual to bulk field $\a$. From the form of our Ansatz (\ref{ansatz}) we know that $\langle \mathcal{O}_\chi\rangle=0$ for our solutions, so the third term in (\ref{Ward}) is generally absent.

\begin{figure}[t!]
\centering
\includegraphics[width=11cm]{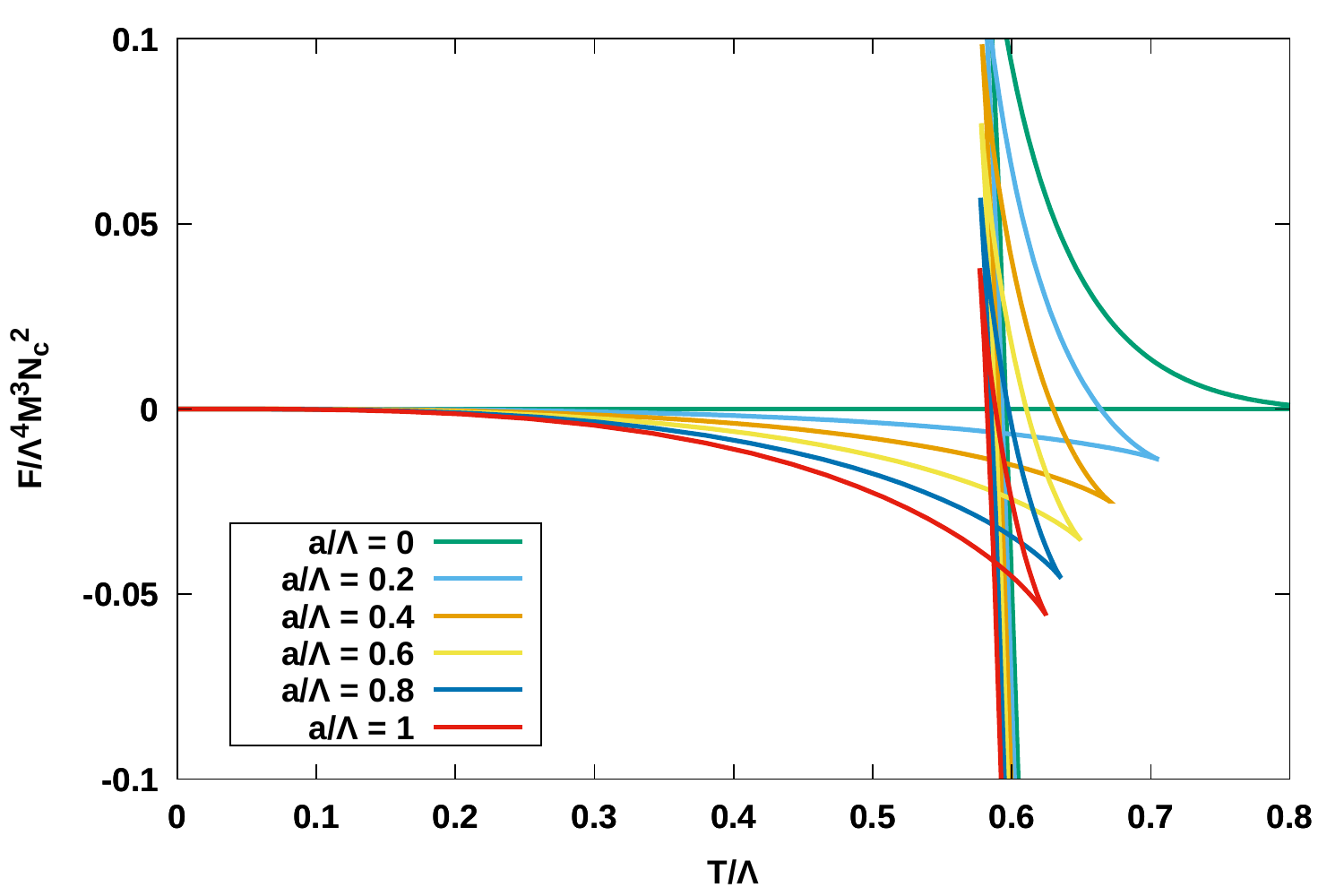}
\begin{picture}(0,0)
\put(-145,90){{\small $I$}}
\put(-98,50){{\small $II$}}
\end{picture}
\caption{\label{fig:freeenergy}Free energy $F/\Lambda^4$ as a function of $T/\Lambda$ for $x=0$ and various values of the anisotropic parameter $a/\Lambda$. The curves develop a characteristic swallow tail behavior at intermediate temperatures, with a first order phase transition connecting two different branches of black hole solutions, $I$ and $II$, which dominate the small $T/\Lambda$ and large $T/\Lambda$ regimes, respectively. At very small temperature, the free energy and entropy density scale as $F\,\propto-T^3$ and $s\,\propto\, T^2$, respectively, while for very large temperature they scale as $F\,\propto-T^4$ and $s\,\propto\, T^3$.  All the thermodynamic quantities jump discontinuously at the transition. Only for $a=0$ the thermal gas solution dominates at small temperature.}
\end{figure}
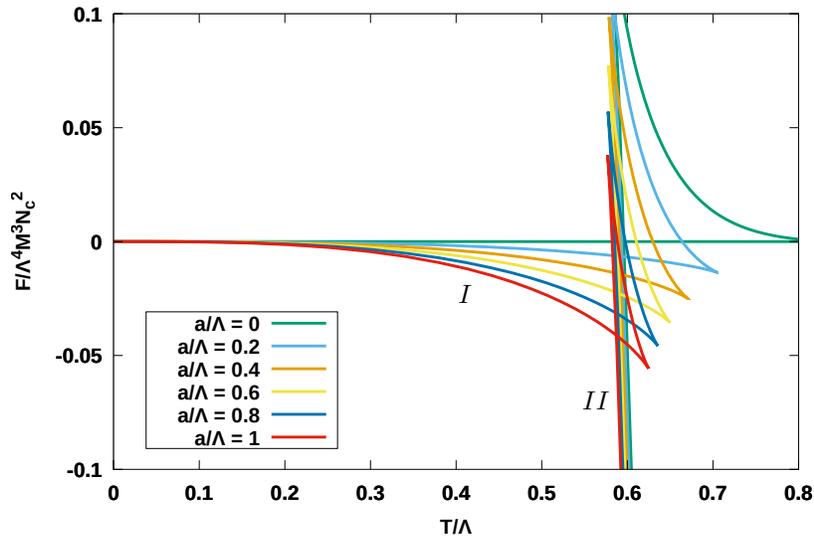

In Figure \ref{fig:freeenergy} we plot the free energy $F/\Lambda^4$ as a function of $T/\Lambda$ for various values of $a/\Lambda$ in the case where $x=0$ so the flavor physics is suppressed. Other values of $x$ behave qualitatively similarly\footnote{As we shall show below, for $x > 0$ there are some additional features which do not affect the main points discussed here.} so we will omit their discussion for simplicity. We observe the following general behavior: i) for small $T/\Lambda$ there are always two solutions.  First, there is the ground  state heated up to temperature $T$, which is referred to as the thermal gas solution. The gravitational background dual to this state is obtained from the black brane solution (\ref{ansatz}) by sending $f(r)\to1$ and compactifying the Euclidean time direction with period $1/T$. We take this solution as our reference background for the free energy computation so it corresponds to the horizontal axis $F=0$ in the figure.\footnote{States dual to black hole solutions have free energies of order $\mathcal{O}(N_c^2)$, while the thermal gas solution has a free energy of order $\mathcal{O}(N_c^0)$.} Second,  we also observe a black hole solution which, in fact, dominates the ensemble at small $T/\Lambda$ (for non-zero $a$). The fact that there is a black hole branch that dominates at small $T/\Lambda$ is in stark contrast with the standard, isotropic models of holographic QCD, for which the thermal gas solution is
the dominant one at low temperatures.\footnote{This is in analogy to the results for the canonical ensemble of charged black holes, where the anisotropy parameter $a$ plays the role of the charge \cite{Chamblin:1999tk,Caceres:2015vsa,Pedraza:2018eey}. The swallow tail behavior in these cases is associated to a Van der Waals liquid-gas phase transition.} Indeed, this behavior at low $T/\Lambda$ is crucially related to the fact that, for any non-zero $a$, the IR structure of the theory is drastically modified by the anisotropic deformation, approaching a nearly AdS$_4$ geometry in the deep IR (with logarithmic corrections). At very low temperatures, $T/\Lambda\ll1$, the free energy scales as $F\,\propto -T^3$ while the entropy density scales as $s\,\propto\, T^2$. ii) The free energy exhibits a swallow tail behavior at intermediate values of $T/\Lambda$. In this range of temperatures, besides the thermal gas solution, there are up to three black hole solutions with free energies of order $\mathcal{O}(N_c^2)$. Two of these solutions are sub-dominant in the ensemble, and the third one dominates in this regime. iii) At large enough temperatures there is only one black hole solution ---in addition to the thermal gas solution--- which dominates the ensemble. Since the geometry is asymptotically AdS$_5$ the free energy and entropy density at very high temperatures, $T/\Lambda\gg1$, scale as $F\,\propto-T^4$ and $s\,\propto\, T^3$, respectively. Altogether, if one follows the dependence of the dominant phase as a function of $T$, one finds a single first order phase transition connecting two different branches of black hole solutions, $I$ and $II$ as shown in Figure \ref{fig:freeenergy}, that dominate at small and large temperatures, respectively. 

In the limit $a \to 0$ the free-energy of the black hole solutions $I$ in figure \ref{fig:freeenergy} apparently vanishes which suggests that they approach the thermal gas solution. This is indeed the case: the horizon is pushed deeper and deeper in the IR as $a$ decreases, and disappears from the limiting solution. Therefore the thermodynamics approaches smoothly that of the standard symmetric ($a=0$) IHQCD~\cite{Gursoy:2008bu,Gursoy:2008za} and the black hole phase $I$ is replaced by the thermal gas solution. This is also consistent with the analysis of the zero temperature IR geometry in the same limit in section~\ref{sec:numericalIR}.

\begin{figure}[t!]
\begin{center}
\includegraphics[width=14cm]{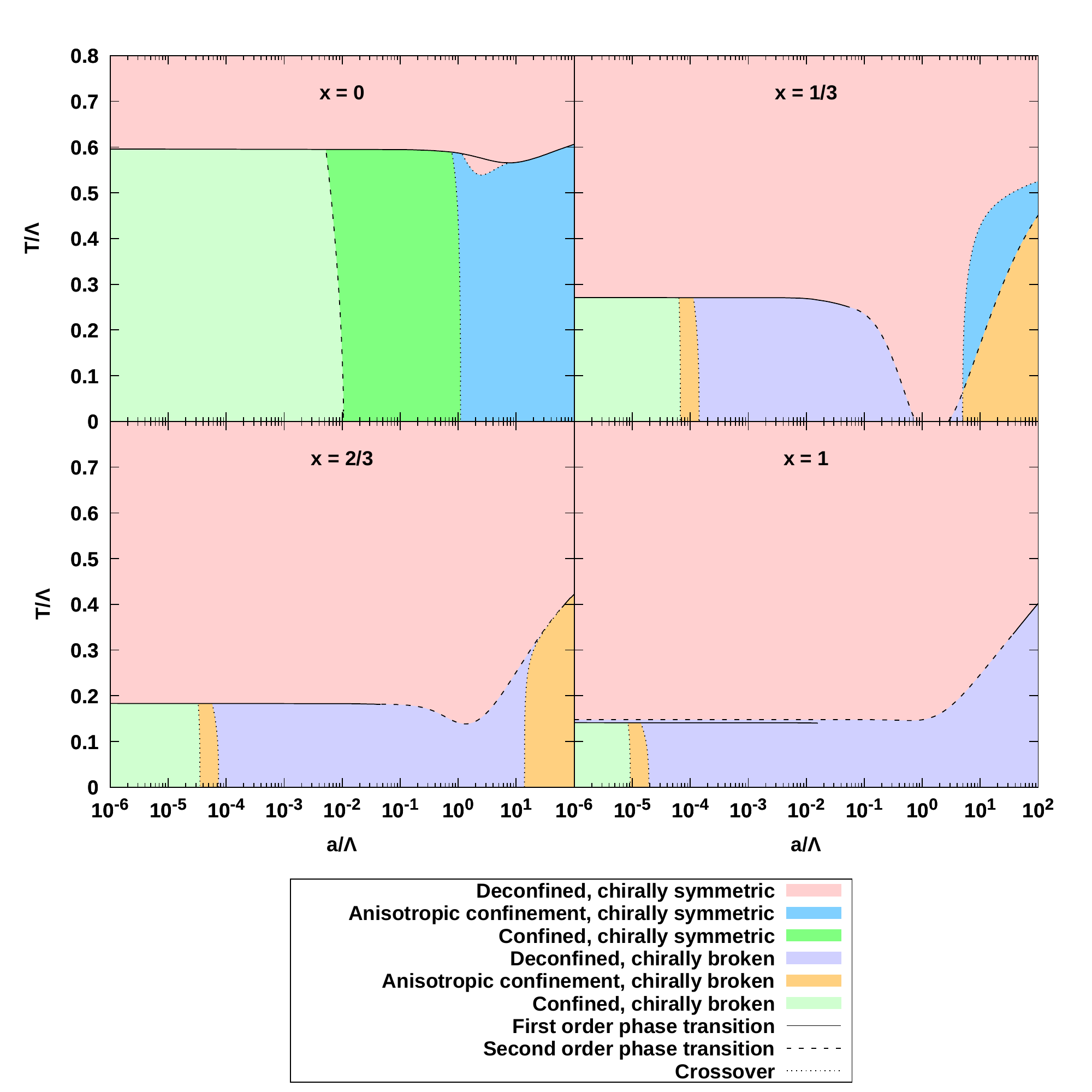}
 \caption{Phase structure of anisotropic holographic QCD in the Veneziano limit, for different values of the ratio $x=N_f/N_c$. We distinguish between different branches of black hole solutions that dominate in the different regimes of $a/\Lambda$ and $T/\Lambda$ (separated by a first order or second order transition), chirally symmetric and chirally broken phases, and confined/deconfined phases as indicated by the behavior of the quark-antiquark potential. See the main text for a detailed explanation.\label{Fig:phases}}
 \end{center}
\end{figure}

In Figure \ref{Fig:phases} we plot phase diagrams for some representative values of $x=N_f/N_c$.
Specifically, the values that we consider are the following: $x=0$, $x=1/3$, $x=2/3$ and $x=1$, respectively. In these diagrams, the black solid lines correspond to a first order transition between two black hole solutions. Such a transition is present for the $x=0$ case (discussed in the previous paragraph) regardless the value of the anisotropic parameter $a/\Lambda$. We notice that for $x=1/3$ and $x=2/3$ this transition eventually becomes second order at large enough anisotropies, while for $x=1$, one can distinguish two transitions, one of first order (which disappears at large anisotropies) and another one of second order.
The second order transitions are shown as black dashed lines. Curiously, the second order line eventually becomes first order again for large values of $a$ only in the case $x=1$.
If one were to plot the analog of figure \ref{fig:freeenergy} for $x > 0$, the swallowtail structure would disappear at the value of $a$ for which the first order transition disappears.
Notice that the results at $a=0$ are consistent with the earlier findings~\cite{Alho:2012mh,Alho:2013hsa,Iatrakis:2014txa,Alho:2015zua,Gursoy:2016ofp,Gursoy:2017wzz} in the model with the same choices for the potentials.

We also distinguish between chirally symmetric and chirally broken phases in figure \ref{Fig:phases}, depending on the value of the quark condensate $\langle \bar{q}q\rangle$ (whether it is zero or non-zero, respectively). For the case $x=0$ the chiral symmetry is determined through the solutions to the DBI action in the probe limit, $x \to 0^+$. See section \ref{sec:chiral} for a more detailed discussion on the quark condensate and the role of the anisotropy. The IR behavior at finite $x$ is generally dominated by the slow rolling tachyonic phase discussed in section \ref{sec:rolling}. The only exception is the $x=1/3$ case, for an intermediate regime of $a/\Lambda$, roughly between $a/\Lambda=1$ and 3. In this range of anisotropies, the low temperature regime realizes the chirally symmetric IR fixed point discussed in section \ref{sec:FP}. Finally, following the criterion discussed in \cite{Gursoy:2007er,Kinar:1998vq}, we indicate whether or not the quark-antiquark potential (for quarks separated in the longitudinal or transverse directions) has a linear behavior, indicating confinement in both directions, anisotropic confinement or deconfinement. These different phases are separated by black dotted lines, indicating a crossover.
We observe that as $x$ is increased, the value of $a$ up to which we have confinement---according to this criterion---decreases. We discuss the explicit calculation of the quark-antiquark potential in section \ref{sec:qqpot}.

An interesting comparison can now be made between figure \ref{Fig:phases} and the phase diagrams of \cite{Gursoy:2016ofp}\@.
In the latter case, the chiral transition temperature first decreases as a function of the magnetic field $B$, and then at very large values of $B$, it starts increasing again.
In figure \ref{Fig:phases}, this behavior is qualitatively similar, with the anisotropy parameter taking the role of the magnetic field.
Note indeed that we see the chiral transition temperature first decrease as a function of $a$, then increase beyond some large enough $a$\@.
This draws an interesting parallel between the effect of a magnetic field on the chiral transition and that of an anisotropy. In the following subsection we make this analogy more concrete by studying the analog of the magnetic susceptibility. Further in section~\ref{sec:chiral} we demonstrate that the chiral condensate also behaves in the same way in the presence of an anisotropy as it does under the influence of magnetic field. In particular, we find an effect analogous to the `inverse magnetic catalysis'~\cite{Bali:2011qj,Bali:2012zg}\@.

\subsection{Anisotropic susceptibility\label{sec:chia}}

\begin{figure}[t!]
\begin{center}
\includegraphics[width=0.6\textwidth]{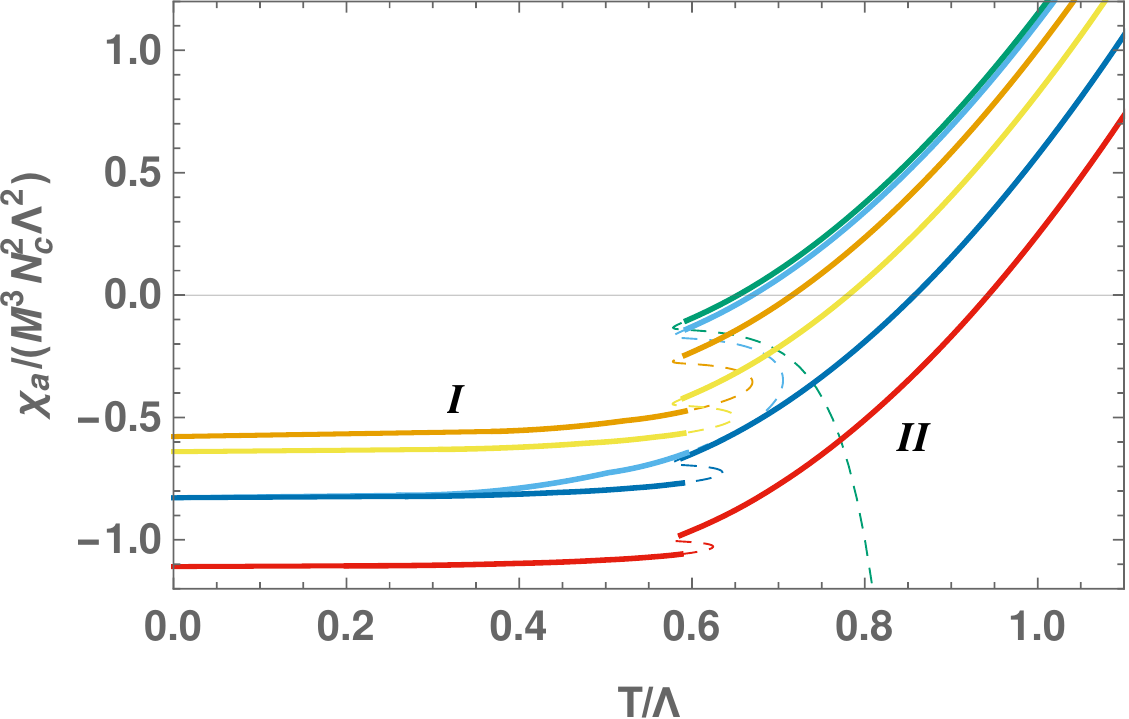}

\vspace{2mm}

\includegraphics[width=0.6\textwidth]{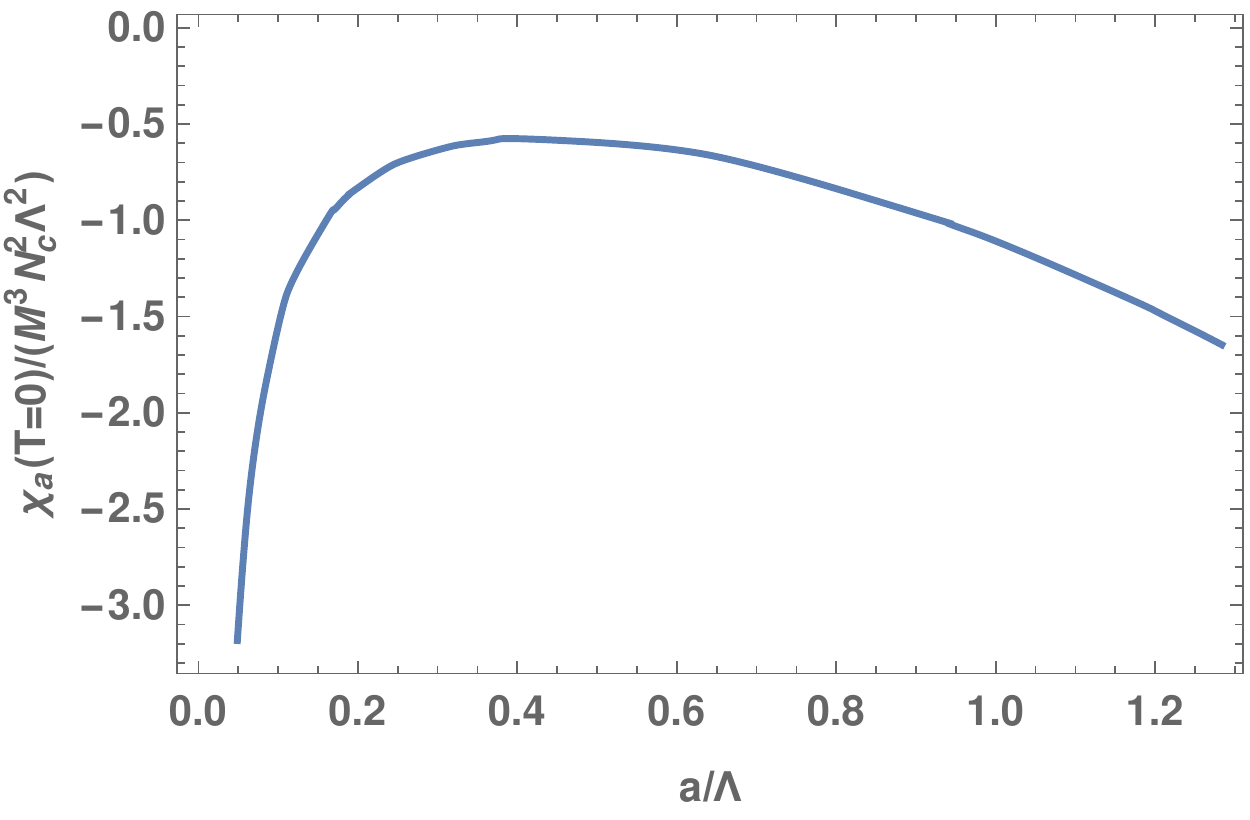}
 \caption{\label{Fig:chia} Top: The anisotropic susceptibility $\chi_a$ as a function of $T/\Lambda$ for $x=0$ and various values of the anisotropic parameter $a/\Lambda$ (the color coding is the same as in Fig.~\protect\ref{fig:freeenergy}). Bottom: $\chi_a$ at zero temperature as a function of $a/\Lambda$.}
 \end{center}
\end{figure}

In analogy to magnetization in a theory with nontrivial external magnetic field, we define the response to the anisotropy parameter $a$ as
\[
M_a = -\frac{\partial F}{\partial a}\ .
\]
In the absence of a better name, we call this object the ``anisotropization''. Let us also
define
\[
\chi_a = \frac{M_a}{a}\ ,
\]
which, at $a = 0$, coincides with the usual definition of a susceptibility.\footnote{This definition is slightly different from that in \cite{Andrade:2013gsa}\@. We are changing the convention to make the analogy to the magnetic field case more explicit.}
Holographically, this `anisotropic susceptibility' can be computed as follows.
Because the free energy is given by the on-shell action, it is enough to consider the derivative of the explicit $a$ dependence in~\eqref{action}. Substituting $\chi = a x_3$ we find that
\be \label{chiintegral}
 \chi_a = - M^3 N_c^2 \int_0^{r_h} dr\, e^{3 A - W} Z(\l) \ .
\ee
This integral is however UV divergent, and needs to be renormalized~\cite{Papadimitriou:2011qb}. We discuss the renormalization procedure in Appendix~\ref{app:renormalization}.
The result obtained by evaluating the renormalized integral numerically for $x=0$ can be seen in Fig.~\ref{Fig:chia}. Notice that $\chi_a$ is not well-defined in the confined phase (I) when $a=0$, as the integral~\eqref{chiintegral} is IR divergent. We see that $\chi_a$ decreases without limit as $a \to 0$ at small temperatures as demonstrated in the bottom plot. Notice that the susceptibility contains a scheme-dependent piece $\propto c_1 + a^2 c_2$ (see Appendix~\ref{app:renormalization}), but the divergence is scheme-independent.

As was done in \cite{Gursoy:2016ofp,Ballon-Bayona:2017dvv}, one can then derive the following relation for the transition temperature of a first order transition as a function of $a$:
\be \label{Tderrel}
\frac{\mathrm{d}T_c}{\mathrm{d}a} = -\frac{a\Delta\chi_a}{\Delta s},
\ee
where $\Delta\chi_a$ and $\Delta s$ are the differences between the two phases of the susceptibility and the entropy density, respectively.
From this equation we can relate the sign of the slope to the jumps of the derivatives of $\chi_a$ and $s$\@. As seen from Fig.~\ref{Fig:chia}, for $x=0$ and $0\le a/\Lambda \le 1$ the jump $\Delta\chi_a$ is positive which agrees with $T_c$ decreasing with $a/\Lambda$ in this region in Fig.~\ref{Fig:phases}\@.
We have checked that at large values of $a/\Lambda$, where $T_c$ increases with $a$ (see Fig.~\ref{Fig:phases}), $\Delta\chi_a$ also has the opposite sign.

A result similar to~\eqref{Tderrel} can be derived for the second order transitions, where $\mathrm{d}T_c/\mathrm{d}a$ depends on the jumps of the first derivatives of $\chi_a$ and $s$ \cite{Gursoy:2017wzz}\@.

\section{Observables\label{sec:obs}}

\subsection{Chiral condensate and inverse anisotropic catalysis\label{sec:chiral}}

\begin{figure}[t!]
\centering
\includegraphics[width=14cm]{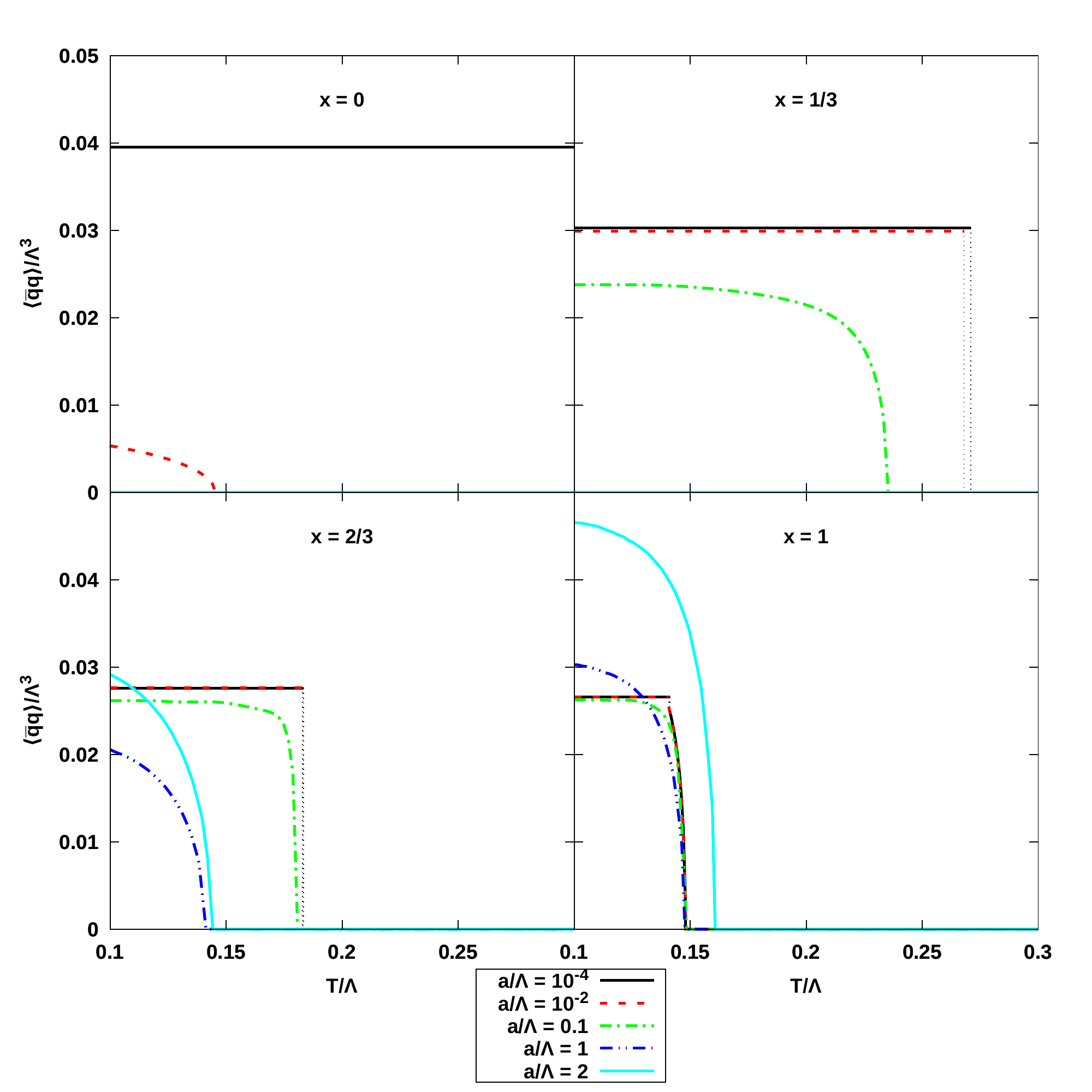}
\caption{\label{fig:qqbar}Chiral condensate $\langle \bar{q}q\rangle/\Lambda^3$ as a function of $T/\Lambda$ for different values of the anisotropy parameter $a/\Lambda$. The four different panels correspond to different values of $x=N_f/N_c$, specifically, to $x=0$, $x=1/3$, $x=2/3$ and $x=1$, respectively. Finite values of $\langle \bar{q}q\rangle$ correspond to a chirally broken phase, as depicted in the phase diagrams of Figure \ref{Fig:phases}. Conversely, a vanishing chiral condensate correspond to a chirally symmetric phase.}
\end{figure}

Once a numerical solution for the metric and other bulk fields is obtained, with boundary conditions as in (\ref{boundaryassymp}), we can extract various observables of interest. In this section we will start by studying the chiral condensate $\langle \bar{q}q\rangle$, which can be computed from the normalizable mode of the tachyon. More specifically, for $m_q=0$, we obtain that the leading behavior of the tachyon near the boundary is given by:
\be
\tau(r)\to\langle \bar{q}q\rangle \,r^3(-\log(r\Lambda))^{\frac{\gamma_0}{b_0}}\,,
\ee
with $b_0\equiv\frac{1}{3}(11-2x)$ and $\gamma_0\equiv\frac{3}{2}$. In Figure \ref{fig:qqbar} we show the dependence of the chiral condensate $\langle \bar{q}q\rangle/\Lambda^3$  on $T/\Lambda$ for various choices of $a/\Lambda$ and $x$. The $x=0$ results here are obtained by solving the tachyon equation in the probe limit.

One can see that for temperatures above the chiral transition the condensate vanishes, and below the transition it is nonzero.
The order of the phase transitions can also be seen by looking at whether the condensate is continuous across the transition.
For $x = 0$, we observe that the chiral condensate decreases with increasing $a$ for all temperatures.
This is indeed consistent with our claim of inverse anisotropic catalysis, because the only effect in this case is due to the backreaction of the axion field,
 which causes anisotropy even at $x=0$. For finite $x$, we observe two effects:
The condensate first decreases with $a$, and then at larger $a$, it increases again.\footnote{This is not visible for $x = 1/3$, because we do not plot large enough values of $a$\@.} The interplay between these two effects arises due to the backreaction of flavors, which tends to smooth out the effect of the anisotropy.

\begin{figure}[ht]
\centering
\includegraphics[width=0.7\textwidth]{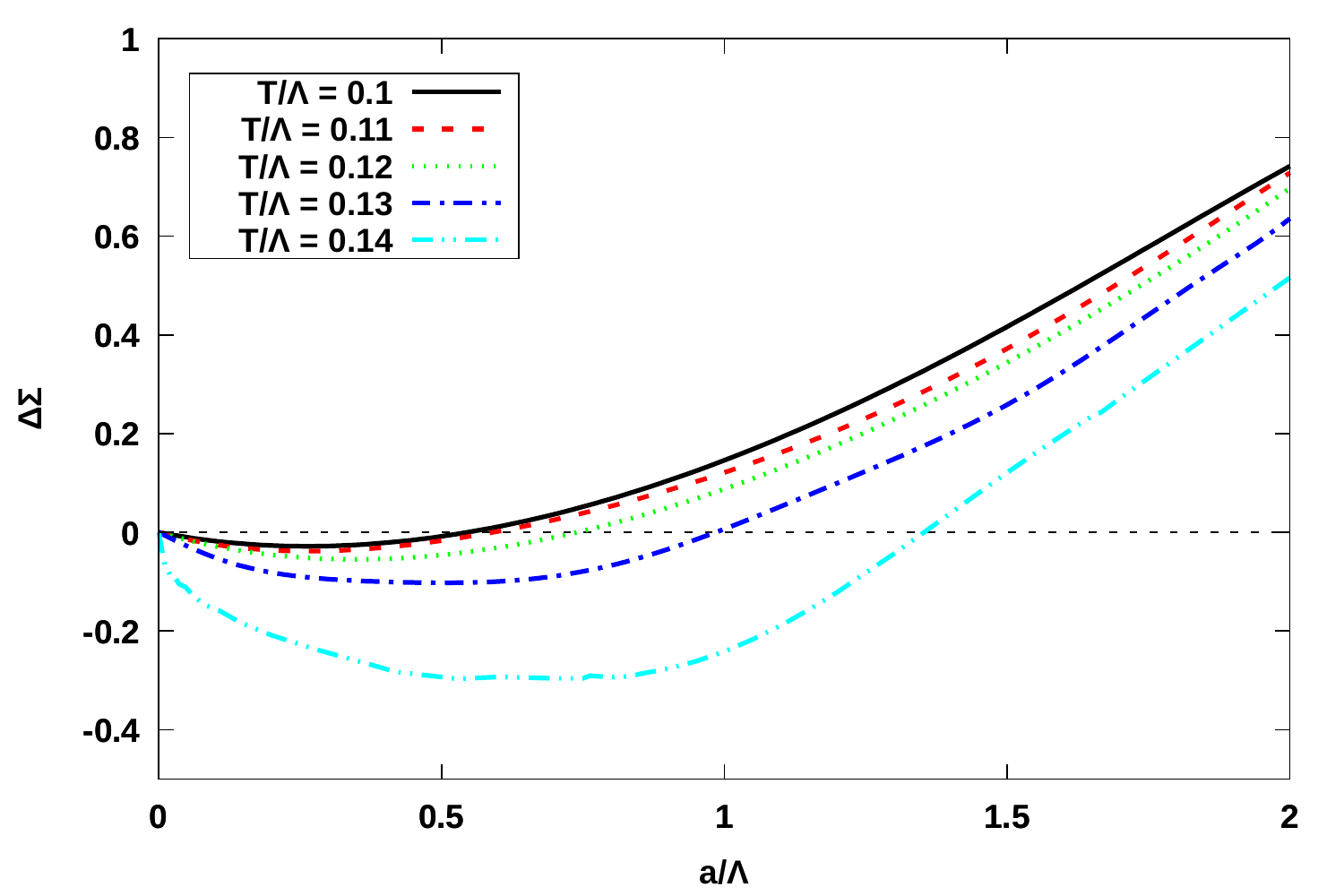}
\caption{\label{fig:deltasigma}$\Delta\Sigma(T,a)$ as a function of $a$ for several constant $T$ values, with $x = N_f/N_c = 1$\@. Note the large slope of the $T/\Lambda = 0.14$ curve near $a/\Lambda = 0$\@. This happens because this temperature is very close to the critical end point of the first order phase transition, which occurs at $a/\Lambda \sim 0.01$\@. Some small numerical noise was removed from this figure by fitting to a polynomial.}
\end{figure}

In order to study this behavior in more detail, it is convenient to define
\[
\Sigma(T,a) = \frac{\langle\bar qq\rangle(T,a)}{\langle\bar qq\rangle(0,0)}, \qquad \Delta\Sigma(T,a) = \Sigma(T,a) - \Sigma(T,0).
\]
This combination is analogous to the quantity which has been computed on the lattice at finite magnetic field (see for example \cite{Bali:2012zg})\@.
In figure \ref{fig:deltasigma}, $\Delta\Sigma$ is plotted as a function of $a$ for different temperatures, and for $x = 1$\@.
We see first a decrease of the condensate with $a$ for small $a$ followed by an increase at larger $a$\@.
The decrease of the condensate is most pronounced for temperatures close to the chiral transition.
This is in close analogy to the analysis of \cite{Gursoy:2016ofp}, which is the same model as in this work, setting $a = 0$, and including a magnetic field in the flavor action.
There, it was found that for small temperatures, the condensate always increases with a magnetic field $B$, while for temperatures close to the chiral transition, there is first a decrease of the condensate with $B$, and then an increase.\footnote{In \cite{Gursoy:2016ofp}, there is also a direct coupling of the magnetic field to the condensate, in addition to the backreaction through the geometry. This direct effect was found to always increase the condensate, in agreement with lattice results \cite{Bruckmann:2013oba}\@. An analogous direct coupling is absent in this work.}
The decrease of the condensate (inverse magnetic catalysis) with a magnetic field was discovered on the lattice \cite{Bali:2011qj,Bali:2012zg}\@.
The analogous behavior we observe here is therefore evidence for the claim made in \cite{Giataganas:2017koz}, namely that a possible cause for the inverse magnetic catalysis is the anisotropy, which can be induced by the magnetic field as in the lattice studies or explicitly as we do in this paper. What we see is that just the presence of anisotropy has the same effect as the magnetic field. We thus call this behavior ``inverse anisotropic catalysis''\@.

\subsection{Particle spectra}\label{sec:spectra}

We observe from the phase diagram, figure \ref{Fig:phases}, that anisotropy may destroy confinement even at relatively low values of $a/\Lambda$. In this section we analyze this phenomenon further by discussing how the meson and glueball states (at zero temperature) melt as $a$ is increased. Dissociation of mesons due to anisotropy has also been studied in other holographic models~\cite{Chernicoff:2012bu,Avila:2016mno}\@.

We will discuss the particle spectrum in the helicity two glueball tower which is relatively easy to analyze.
We have also computed that the spectral functions in other sectors, including all flavored states, and helicity one flavor singlet states, and find a similar behavior to that of the helicity two glueballs. The fluctuations in these additional sectors are presented in Appendix~\ref{app:flucts}.

In order to analyze the helicity two glueballs we write an Ansatz for the perturbation of the metric as
\be
 \delta g_{12} = \delta g_{21} = e^{2 A(r) }e^{ik_\mu x^\mu}h(r)
\ee
where the sum in the plane wave term goes over the time and space indices.  By boosting in the $x_{1,2}$ directions (and assuming that $|\omega|>|q_\perp|$) we can transform the momentum to the form $q^\mu = (\omega,0,0,q)$. We find the following equation for the fluctuations in this frame:
\be \label{h12eq}
h''(r) +3 A'(r) h'(r) + W'(r)h'(r) - q^2 e^{-2 W(r)} h(r) +\omega ^2 h(r) = 0 \ .
\ee
The Schrodinger form is obtained by defining $h(r) = e^{-3A(r)/2-W(r)/2} \psi(r)$:
\begin{align}
 &-\psi''(r) + V_s(r) \psi(r) = (\omega ^2- q^2 e^{-2 W(r)}  )\psi(r) &\\
 &V_s(r) = \frac{1}{2} \left(3 A''(r)+W''(r)\right)+\frac{1}{4}\left(3 A'(r)+W'(r)\right)^2 \ . &
\end{align}
Inserting here the asymptotic IR solution~\eqref{IRasympt} from above, i.e., $\text{AdS}_4 \times \mathbb{R}$ with logarithmic corrections, we find that $V_S(r) \sim 2/r^2$ as $r \to \infty$, indicating that the spectrum is continuous. That is, the glueballs have finite widths even at zero temperature, signaling the possibility of the decay to the $\text{AdS}_4$ vacuum.

Spectral density of the helicity two glueballs can be obtained from the correlator of the energy momentum tensor $T_{12}$, which is determined (in momentum space) through the UV coefficient of the IR-regular solution to~\eqref{h12eq}. The UV expansion of the properly normalized solution is given by
\be \label{hUVexp}
 h(r) = 1 + \mathcal{O}(r^2) + G(\omega,q) r^4\left[1+\mathcal{O}\left(\frac{1}{\log r}\right)\right]
\ee
where $G(\omega,q)$ is the correlator. For positive $\omega$, definition of IR regularity is obtained via analytic continuation from the upper half of the complex $\omega$-plane which picks up the solution with the IR asymptotics $\psi(r) \,\propto\, e^{i\omega r}$ (i.e. the sign in the exponent is plus) so that $h(r)\, \propto\, e^{-3A(r)/2-W(r)/2} e^{i\omega r}$. Notice moreover that the higher order corrections to the source term in~\eqref{hUVexp} are real for real $\omega$. This is the case because then the fluctuation equation has real coefficients, and the only source for complex behavior is the IR boundary condition which only affects the integration constants (i.e., the correlator) in the UV expansion. Therefore
\be
 \mathrm{Im}\, h(r) = \mathrm{Im}\, G(\omega,q)\, r^4\left[1+\mathcal{O}\left(\frac{1}{\log r}\right)\right] \ , \qquad (\omega>0)
\ee
for the solution with the normalization $h \to 1$ in the UV. The coefficient of the $r^4$ term here is the spectral density which can be therefore extracted unambiguously.
\begin{figure}[t!]
\centering
\includegraphics[width=11cm]{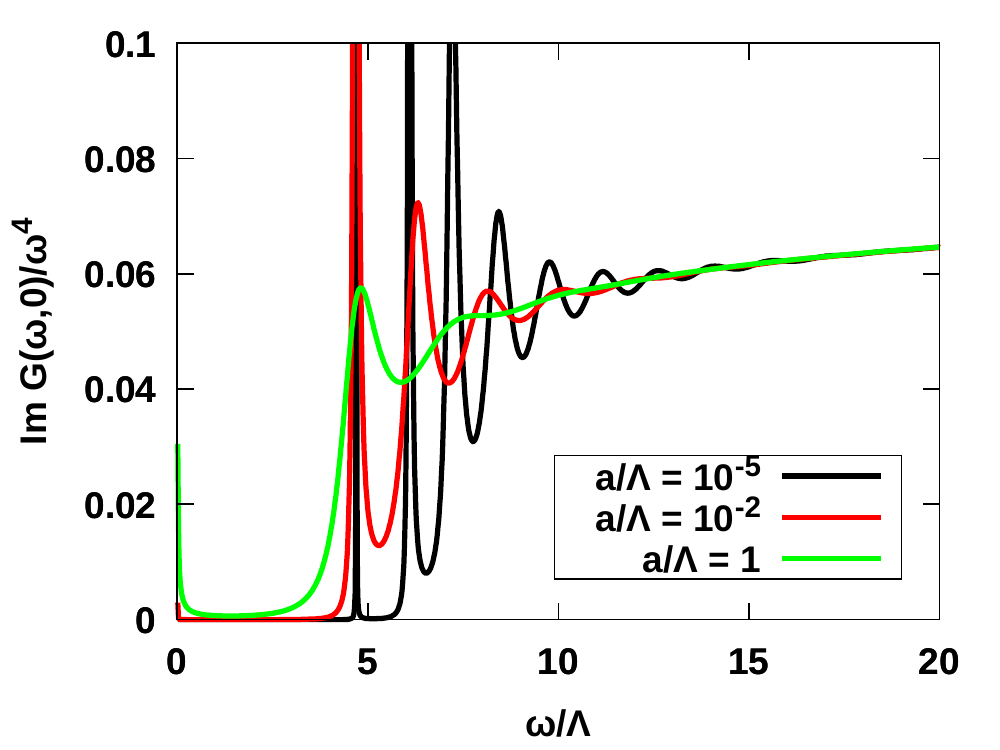}
\caption{\label{fig:glueballspectrum}The spectral function in the helicity two glueball sector at $x=0$ for various values of $a$.}
\end{figure}
\begin{figure}[t!]
\centering
\includegraphics[width=11cm]{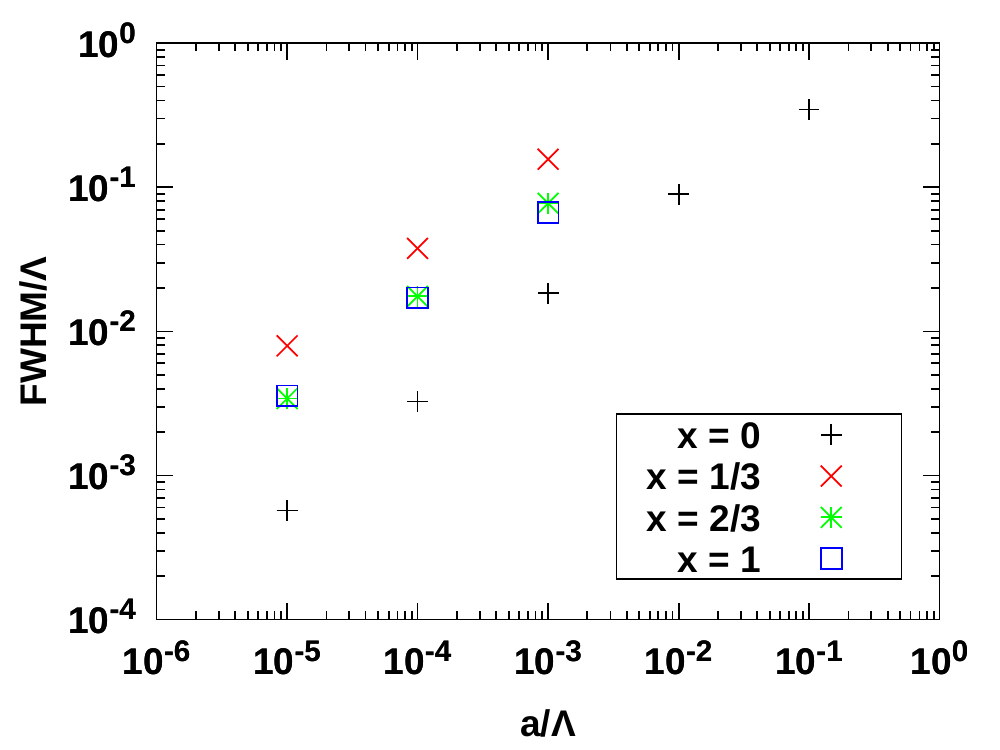}
\caption{\label{fig:widths}The widths of the lowest helicity two glueball states as a function of $a$ for various values of $a$. The plotted width is the full width at half maximum (FWHM) of the peak.}
\end{figure}

We show the resulting spectral density, normalized by $\omega^4$, for $x=0$ and for three choices for the values of $a$ in Fig.~\ref{fig:glueballspectrum}. As expected, the glueballs have finite widths even though we are working at zero temperature. At $a/\Lambda = 10^{-5}$ the first 3-4 states are clearly peaked. As $a$ is increased, the widths of the glueballs grow, and for $a/\Lambda=1$ the lowest glueball is already very wide while the other peaks have melted away. Despite this the spectral density is still heavily suppressed in the regime corresponding to the mass gap at $a=0$, i.e., for $0<\omega/\Lambda\lesssim 4$. We also note that the peak arising at $\omega = 0$ is due to the IR behavior $\mathrm{Im}\,G(\omega,0)\, \propto\, \omega^3$ which reflects the IR geometry of the system being approximately AdS$_4$.

For higher values of $x$ the spectral density is qualitatively similar to $x=0$. We show the dependence of the width (more precisely the full width at half maximum) of the lowest glueball mode on $a$ for $x=0$, $1/3$, $2/3$ and $1$ in Fig.~\ref{fig:widths}. The widths first increase when $x$ grows from $0$ to $1/3$ but then decrease as $x$ is further increased.

\subsection{Quark-antiquark potential\label{sec:qqpot}}

In the absence of anisotropy in the low-temperature phase, our model is known to show linear confinement, i.e.~the quark-antiquark potential $V$ grows linearly with the separation $L$ of the test quark and the antiquark for large enough $L$\@.
A potential of such a shape prevents colored particles from escaping their bound states and becoming free.
On the other hand, in section \ref{sec:spectra} the glueball states were shown to melt at zero temperature in the presence of anisotropy indicating that quarks can in fact escape their bound states in an anisotropic vacuum state. In this section, we investigate the mechanism behind this decay in some more detail by studying quark-antiquark potential at zero temperature. See also~\cite{Arefeva:2018hyo,Arefeva:2018cli} for the analysis of the potentials in an anisotropic setup in slightly simpler models.

In holography, the quark-antiquark potential is a sum over saddle points, which has several terms contributing \cite{Bak:2007fk}\@.
In principle, one should take all of these into account, but as a first approximation, we work in the $\alpha' \rightarrow 0$ limit, where only the one with the smallest action contributes.
Also, we choose to neglect the graviton exchange contribution that plays an important role by maintaining smoothness of the Polyakov loop two-point function in $L$ \cite{Bak:2007fk}, because this contribution does not have a qualitative effect on our results.
The remaining terms can be computed by evaluating the on-shell Nambu-Goto action for a static string in the 5D spacetime described by the string frame metric, with the endpoints of the string attached to the AdS boundary \cite{Maldacena:1998im,Rey:1998ik}.
Following \cite{Gursoy:2007er,Kinar:1998vq}, we find that at large $L$, the quark-antiquark potential parallel to the anisotropy $V_\parallel$ grows linearly with $L$ if $A_S + W/2$ has a minimum, with $A_S = A + \frac{2}{3}\log\lambda$ the string frame scale factor.
Similarly, we find that the quark-antiquark potential perpendicular to the anisotropy $V_\perp$ grows linearly with $L$ at large $L$ if $A_S$ has a minimum.
In principle, one could also look at the potential in arbitrary directions, but for lack of symmetry this is more difficult, and will be left for future work.

The presence of a minimum in $A_S$ resp.~$A_S + W/2$ described above is the criterion used to label a phase as confining in the phase diagram (Fig.~\ref{Fig:phases})\@.
However, while this indeed gives some indication of where in the phase diagram there is confinement, it does not give the full picture.
This can be seen in figure \ref{fig:glueballspectrum}, where we see that the first glueball is already no longer absolutely stable in regions that satisfy the confinement criterion used in figure \ref{Fig:phases}\@.
In figure \ref{fig:glueballspectrum}, we also see that the first glueball peak is still very narrow for anisotropies orders of magnitude larger than the anisotropy for which the minimum of $A_S (+ W/2)$ disappears.
To address this discrepancy, one has to compute not just the large $L$ behavior of $V$, but the entire function.

\begin{figure}[t!]
\centering
\includegraphics[width=14cm]{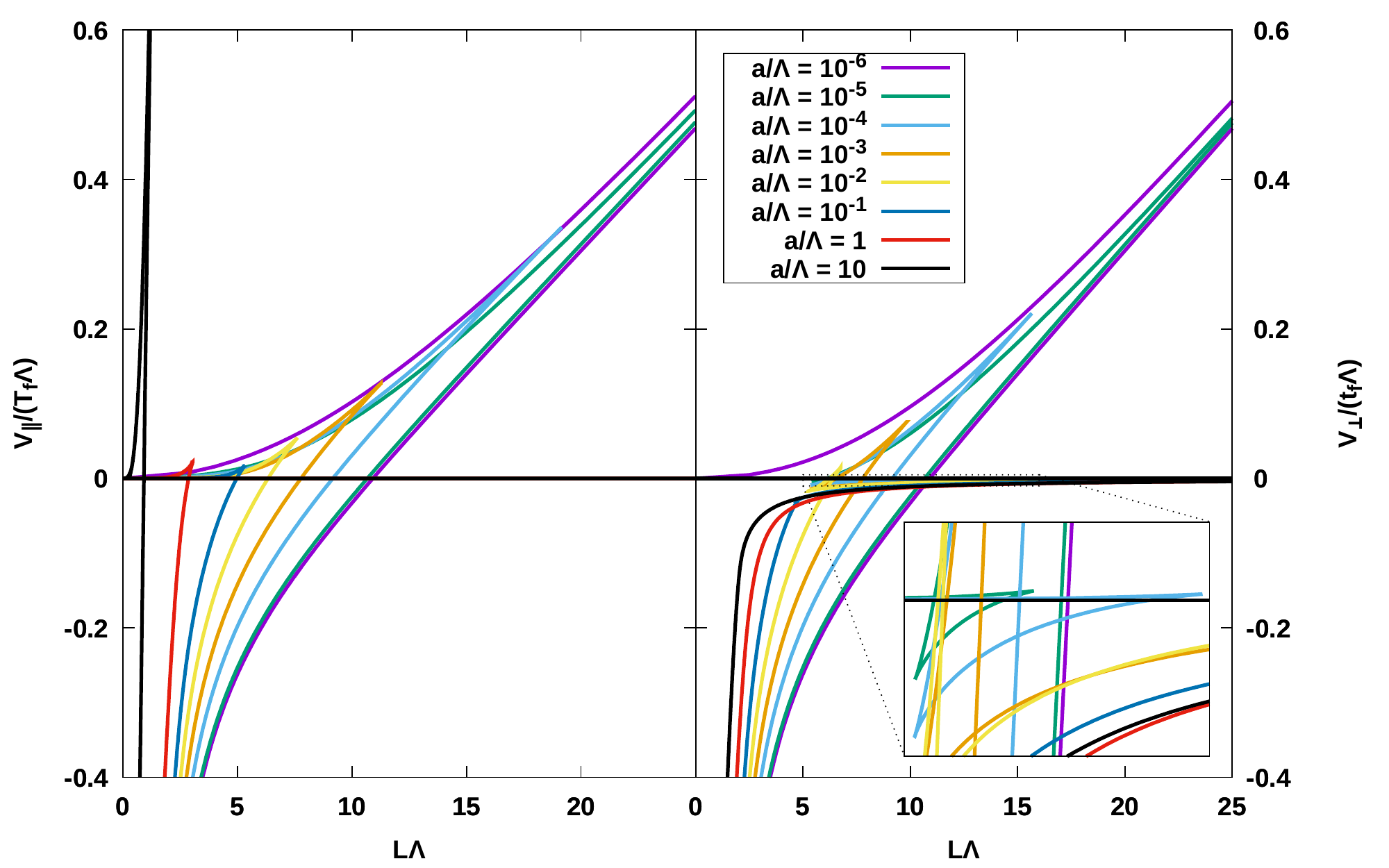}
\caption{\label{fig:qqpot}Quark-antiquark potential in the direction parallel to the anisotropy $V_\parallel$, and in the direction perpendicular to the anisotropy $V_\perp$, for $x = 1$\@. Note that each curve has multiple branches, and that every curve has a branch which is identically zero. Also, for $a/\Lambda = 10^{-6},10^{-5}$, the linear branch of $V_\parallel$ continues to infinity, while for $V_\perp$ this only happens for $a/\Lambda = 10^{-6}$\@. Note that not just the branch with the smallest action is plotted. In the approximation we   consider the quark-antiquark potential at separation $L$ is given by the lowest branch.}
\end{figure}

Instead of computing $V_\parallel$ as a function of $L$, we can obtain both $V_\parallel$ and $L$ as functions of the worldsheet turning point in the bulk $r_F$ in the following way \cite{Gursoy:2007er, Kinar:1998vq}:
\begin{align} \label{Vparallel}
\frac{V_\parallel(r_F)}{T_f} &= e^{2A_S(r_F)+W(r_F)}L(r_F) + 2\int_0^{r_F}\frac{\mathrm{d}r}{e^{W(r)}}\sqrt{e^{4A_S(r)+2W(r)} - e^{4A_S(r_F)+2W(r_F)}}&\nonumber\\
& \phantom{=} - 2 \int_0^{\infty}\mathrm{d}r\, e^{2 A_S(r)}\ ,&\\
L(r_F) &= 2\int_0^{r_F}\frac{\mathrm{d}r}{e^{W(r)}}\frac{1}{\sqrt{e^{4A_S(r)+2W(r)-4A_S(r_F)-2W(r_F)} - 1}}\ ,&\label{eq:VparL}
\end{align}
with $T_f$ the string tension. The last term in~\eqref{Vparallel} is the UV regulator which in our convention equals (twice) the action of a straight string hanging from the boundary down to the IR.
Similarly, $V_\perp$ can be obtained using the same formulas with $W$ set to 0\@.
Results of these computations for $x = 1$ are shown in figure \ref{fig:qqpot}\@.

One immediate observation is that $V$ has multiple branches.
For each $L$, the smallest $V$ corresponds to the globally stable branch.
Note also that there always exists a zero branch, corresponding to two detached portions of string which have fallen into the deep IR \cite{Bak:2007fk}\@.
The existence of this zero branch is due to the IR behavior which is, as we have seen, qualitatively different from the case without anisotropy.
In the isotropic case the string falling into the deep IR has a diverging on-shell NG action, whereas in the anisotropic case the on-shell action is zero.\footnote{It is exactly zero because of the way the on-shell action is renormalized. This renormalization has no physical effect.}
The existence of the zero branch is important, because it means that even though for small enough nonzero anisotropy $A_S$ has a minimum, indicating linear confinement according to \cite{Gursoy:2007er}, the branch of solutions corresponding to this case may be globally  unstable, meaning that bound states can decay.
We remark that our potentials at nonzero $a$ are similar to those contructed in part by deep learning methods in~\cite{Hashimoto:2018bnb}, but the regime where the potential is linear is much shorter.

One can now also explain why the first glueball peak is still very narrow at anisotropies for which $A_S$ does not exhibit a minimum.
The lowest lying bound state decaying to free particles corresponds to the ends of the string starting close together and ending up freely moving a large distance apart, so that the zero branch is the stable one.
In almost\footnote{For $x = 1/3$ the nontrivial structure of $V$ never goes away, but glueballs still melt. Note that this is also the value of $x$ for which we have the different structure in the phase diagram.} all cases we examined, the melting of the first glueball peak occurs when the nontrivial branches of $V_\perp$ disappear.
Therefore we conjecture that the presence of nontrivial structure makes the worldsheet corresponding to glueball decay have a large action $S_\text{decay}$ associated to crossing this nontrivial structure. In other words, the nontrivial structure in $V_\perp$ prevents the decay of the first glueball.
This would then contribute to a lifetime $\tau \,\propto\, \exp(S_\text{decay})$, which, if $S_\text{decay}$ is large enough, will make the excitations very long-lived.
To support this conjecture, we have to find the corresponding time-dependent worldsheet solutions. This is a hard problem and we leave it for future work.

Finally, we would like to comment that the computation described above assumes that the string lies in the $(X^r,X^3)$-plane when computing $V_\parallel$, and in the $(X^r,X^1)$-plane when computing $V_\perp$.
In principle, while these solutions indeed solve the equations of motion, they could be unstable in the neglected directions. This would lead to yet more branches of solutions.
Again, we leave exploration of these possibilities to future.

\subsection{Entanglement entropy\label{sec:entanglemententropy}}

Another interesting observable probing the IR structure of the zero temperature geometry is the entanglement entropy.
The quark-antiquark potential is determined by a minimal length of a string stretching in the bulk, where the length is computed in the string frame, whereas the entanglement entropy arises from a similar minimization procedure for a higher dimensional surface in the Einstein frame \cite{Ryu:2006bv}.
We compute the entanglement entropy for two different regions of the boundary:
\begin{itemize}
\item A region defined by $0 < x_3 < L$\@.
Note that the $x_3$ direction is parallel to the direction of anisotropy.
We therefore denote the entanglement entropy of this region by $S_{E,\parallel}$\@.
\item A region defined by $0 < x_1 < L$\@.
Since $x_1$ is perpendicular to the direction of anisotropy, we denote the entanglement entropy of this region by $S_{E,\perp}$\@.
\end{itemize}
The problem of determining the entanglement entropy reduces to finding the minimal area of a three dimensional spatial surface in the Einstein frame metric stretching between $x_3 = 0$ and $x_3 = L$ for the first region, and between $x_1 = 0$ and $x_1 = L$ for the second region.
Subtracting the UV divergence in the same way as was done for the quark-antiquark potential, one finds both $S_{E,\parallel}$ and $L$ as functions of the turning point in the bulk $r_F$:\footnote{This regularization defines the entanglement entropy of the whole boundary as 0, i.e.~$S_{E,\parallel}|_{L = \infty} = S_{E,\perp}|_{L = \infty} = 0$.}
\begin{align}
\frac{S_{E,\parallel}(r_F)}{4\pi A_VM^3N_c^2} &= e^{3A(r_F)+W(r_F)}L(r_F) + 2\int_0^{r_F}\frac{\mathrm{d}r}{e^{W(r)}}\sqrt{e^{6A(r)+2W(r)} - e^{6A(r_F)+2W(r_F)}}& \nonumber\\
&\phantom{=} - 2\int_0^{\infty}\mathrm{d}r \,e^{3A(r)}\ , &
\label{eq:EE}\\
L(r_F) &= 2\int_0^{r_F}\frac{\mathrm{d}r}{e^{W(r)}}\frac{1}{\sqrt{e^{6A(r)+2W(r)-6A(r_F)-2W(r_F)} - 1}}\ ,& \label{eq:EE2}
\end{align}
with $A_V$ an infinite factor that arises because the region is spatially infinite in two dimensions.
The analogous formulas for $S_{E,\perp}$
read
\begin{align}
\frac{S_{E,\perp}(r_F)}{4\pi A_VM^3N_c^2} &= e^{3A(r_F)+W(r_F)}L(r_F) + 2\int_0^{r_F}\mathrm{d}r\,\sqrt{e^{6A(r)+2W(r)} - e^{6A(r_F)+2W(r_F)}}& \nonumber\\
&\phantom{=} - 2\int_0^{\infty}\mathrm{d}r \,e^{3A(r)+W(r)}\ , &
\label{eq:EEperp}\\
L(r_F) &= 2\int_0^{r_F}\mathrm{d}r\,\frac{1}{\sqrt{e^{6A(r)+2W(r)-6A(r_F)-2W(r_F)} - 1}}\ .& \label{eq:EEperp2}
\end{align}

\begin{figure}[ht]
\centering
\includegraphics[width=0.8\textwidth]{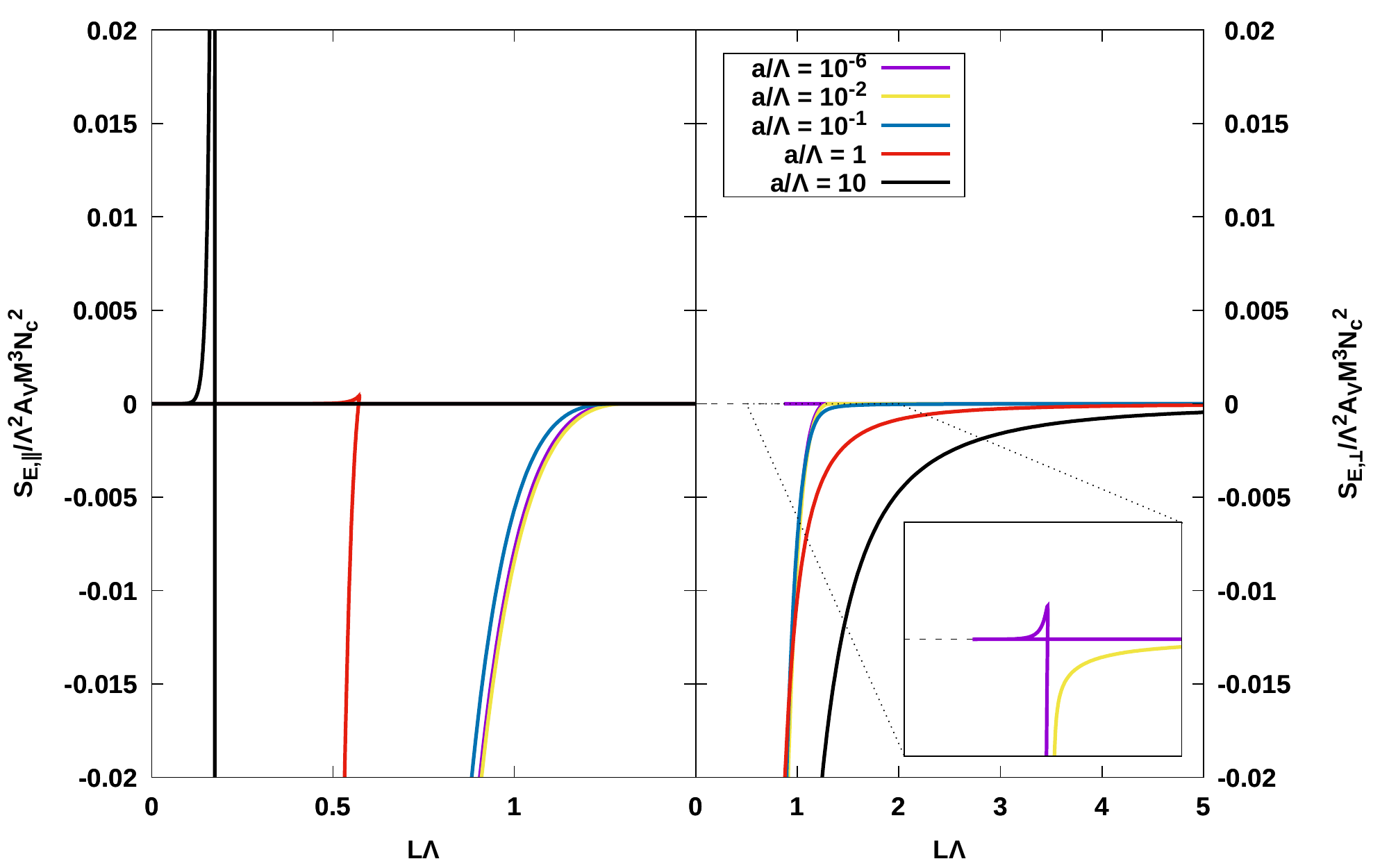}
\caption{\label{fig:entanglemententropy}Entanglement entropy $S_{E,\parallel}$ and $S_{E,\perp}$ for the respective regions described in section \ref{sec:entanglemententropy}, for $x = 1$ and different values of $a$\@. Note that each curve has multiple branches.}
\end{figure}

The result of this computation for $x = 1$ is shown in figure \ref{fig:entanglemententropy}\@. 
Note that in the figure also the unstable branches are shown.
The entanglement entropy corresponds to the lowest branch.

One can see that the results for values of $a/\Lambda$ up to about $0.1$ are similar to isotropic results (see, e.g.,~\cite{Klebanov:2007ws,Dudal:2018ztm})\@.
When the RT surface probes the deep IR, so near where in our normalization the entanglement entropy vanishes, the result depends heavily on the orientation of the entangling region even small values of $a/\Lambda$.
In particular, for $S_{E,\perp}$, the curve of the connected surfaces always reaches $(L = 0, S_{E,\perp} = 0)$, where it connects to the branch of the disconnected solution which have $S_{E,\perp} = 0$ for all values of $L$ in our normalization.
In the case of $S_{E,\parallel}$, however, the point $L=0 = S_{E,\parallel}$ is not reached for any nonzero values of $a/\Lambda$.
The swallowtail structure present at zero anisotropy quickly gets smaller, and eventually vanishes entirely as $a/\Lambda$ grows.

When $a/\Lambda$ is roughly $\mathcal{O}(1)$ the result starts being very different depending on the orientation of the region.
For $S_{E,\parallel}$, the entanglement entropy crosses zero for a smaller value of $L$, while $S_{E,\perp}$ has the opposite behavior. The former behavior is similar to what has been observed in the presence of a magnetic field in a simpler setup~\cite{Dudal:2016joz}\@.
Note that the curves at large $a/\Lambda$ look qualitatively similar to what happens with the quark-antiquark potential.

The most striking difference between the entanglement entropy and the quark-antiquark potential is that the entanglement entropy is almost independent of $a$ for small enough $a$, while the quark-antiquark potential is very sensitive to even small values of $a$\@.
We can explain this difference with the observation that at small $a$, the combination $2A_S + W$ appearing in \eqref{eq:VparL} either has a minimum, or is close to having one (i.e.~$2A_S + W$ is almost flat)\@.
Because of this the result is very sensitive to the boundary conditions of the string and tiny details of the geometry, leading to the observed strong dependence on $a/\Lambda$ even for $a/\Lambda \ll 1$\@. Moreover, we note that the potentials have nontrivial structure up to relatively large value of $L \sim 10/\Lambda$ for tiny values of $a/\Lambda$, demonstrating that the string indeed probes the deep IR geometry.

The combination $3A + W$ (which appears instead of $2A_S + W$) for the entanglement entropies in \eqref{eq:EE2} and $\eqref{eq:EEperp2}$ though, is much steeper. We notice that the nontrivial behavior for the entanglement entropy takes place at smaller values of $L\Lambda$ as compared to the quark-antiquark potential. For the entanglement entropy the characteristic scale is $L \sim 1/\Lambda$, which is to be expected in the absence of nontrivial dynamics and other scales. Therefore the characteristic RT surface remains relatively close to the boundary, $r \lesssim 1/\Lambda$, and is insensitive to the modified IR geometry due to small amounts of anisotropy (see Sec.~\ref{sec:numericalIR} for the discussion on the geometry).

At first glance, one would suspect that this computation suffers from the same potential caveat as the quark-antiquark potential, namely the possibility that the string is unstable towards twisting in the bulk.
However, since in this case the `string' is just a slice of the Ryu-Takayanagi surface, such an instability cannot occur, because it would always amount to an RT surface diffeomorphism.

\section{Conclusions and outlook\label{sec:con}}

In this article we carried out a detailed analysis of anisotropic effects in improved holographic QCD including a fully backreacted quark sector in the Veneziano limit (V-QCD). The anisotropy was sourced by an axion profile with a linear dependence on one of the spatial coordinates, which is equivalent to deforming the theory by a space-dependent $\theta$ term. In turn, the action and the solution still remain translationally invariant along the relevant direction. We found numerical solutions of the equations of motion representing a family of anisotropic black branes and a thermal gas solution that is obtained from the black branes by sending the horizon to the deep interior of the geometry.

The latter thermal gas solution corresponds to the vacuum (zero temperature and entropy), heated up to a temperature $T$, of the corresponding anisotropic gauge theory at strong coupling. First we studied the RG flow in this vacuum state and showed that the IR end point is a scale invariant (almost) fixed point with approximate conformal symmetry $SO(2,3)\times \mathbb{R}$. In the bulk picture this  corresponds to a deep interior geometry $\text{AdS}_4 \times  \mathbb{R}$ up to logarithmic corrections. This behavior is very similar to the IR theory obtained by deforming ${\cal N}=4$ super Yang-Mills with a magnetic field at strong coupling \cite{DHoker:2009mmn, DHoker:2012rlj} which results in $\text{AdS}_3 \times \mathbb{R}^2$ in the bulk. In that case the understanding in the field theory was that the fermions could effectively move only along the $B$ direction at very low energies, due to Landau quantization in the transverse directions, and this gives rise to a $\text{CFT}_2$ which explains $\text{AdS}_3$ in the bulk. In the present case, we may try a similar understanding that directly follows from the picture\footnote{It should be noted however that our holographic model does not precisely correspond to this field theory. This is clear from the fact that the well-defined external symmetry (\ref{sym1}) should correspond to a bulk dual written in terms of $\tilde{A}_5 + d \chi$ where  $\tilde{A}_5$ is a bulk gauge field dual to the axial U(1). We indeed have room for this in the V-QCD theory but decided not to include the dynamics of $\tilde{A}_5$ for simplicity.} of the partition function (\ref{qqint3}) and (\ref{Da}). Modulo coupling to the non-Abelian gauge fields, precisely this field theory is used to study the Weyl semimetals at weak coupling \cite{Grushin:2012mt, Zyuzin:2012tv, Goswami:2012db}. On the other hand, it is quite reasonable to expect that integrating out the gauge fields produces an effective mass term $M \bar{q} q$ for the Weyl quarks, just like in the NJL models. For values of $M> a/c_a$ the Weyl semimetals are in the insulator phase as momentum in the $x_3$-direction becomes gapped \cite{Grushin:2012mt, Zyuzin:2012tv, Goswami:2012db}. This then would restrict the motion of the quarks to the transverse plane and in the far IR would give rise to an approximate $\text{CFT}_3$ which would explain the $\text{AdS}_4$ factor.
This interesting IR fixed point determines many of the interesting aspects we observe in this paper.  We would like to remark that this IR geometry is present even for an infinitesimally small $a$ and is drastically different than the $a=0$ isotropic case. This drastic change was found to be due to a subtle competition between the potentials $V_g$ (which determines the behavior of the pure glue geometry) and $Z$ (which controls the effect of the axion), and leads to a number of geometrical imprints in the IR.

Next we studied the thermodynamics and the phase diagram at finite temperature and found a rich structure with competition between confined/deconfined as well as chirally broken/symmetric phases as a function of $T$ and $a$. This is summarized in figure \ref{Fig:phases}. In particular we found that anisotropy can both deconfine the theory and destroy the condensate depending on the choice of $x= N_f/N_c$. Finally, we calculated  certain observables of interest, such as the chiral condensate, particle spectra, quark-antiquark potentials and entanglement entropy which help investigate the effect of anisotropy both in the vacuum and the plasma states.

Perhaps our most important result concerns the behavior of the chiral condensate as a function of anisotropy. In \cite{Gursoy:2016ofp}, it was conjectured that the backreaction of a magnetic field onto the geometry was responsible for inverse magnetic catalysis, while the direct coupling was responsible for magnetic catalysis.
In the present work we consider an electric neutral plasma, yet observe the same effect of inverse catalysis, due to $a$ instead of $B$. This phenomenon is present even for $x=0$, where there is no backreaction of flavors. The reason is that the anisotropy in this case is generated in the color sector, contrary to the case of the magnetic field. Hence, we dubbed this phenomenon as ``inverse anisotropic catalysis''. At finite $x$ we observe a competition between inverse catalysis and normal catalysis as we vary the temperature $T$, with a very similar pattern than the one produced by a magnetic field. In particular both in the behavior of the chiral transition temperature, as well as the chiral condensate, the similarity is striking. This indicates that the backreaction of flavors tends to cancel out the effects of the anisotropy. These observations
lead us to the conclusion that the usual inverse magnetic catalysis that is observed in the presence of a magnetic field may equally well be caused by the anisotropy brought in the system by it. This is remarkable because, if we extrapolate it to QCD, it would mean that inverse magnetic catalysis could be recast entirely in terms of ``inverse anisotropic catalysis''. It would be extremely interesting to directly check this proposal on the lattice; see more on this below.

Another interesting result in the same context was the identification of a universal order parameter for (inverse) anisotropic catalysis. In order to do so we defined the quantity ``anisotropic susceptibility'' in analogy to the magnetic susceptibility. This susceptibility was then identified as a natural order parameter, similar to what was done in~\cite{Gursoy:2016ofp,Ballon-Bayona:2017dvv} for the magnetic case. We then obtained a relation between the critical temperature of the first order transition and the anisotropic susceptibility in Eq.~\eqref{Tderrel}\@.

One (possibly alarming) property of the IR geometry was the divergence of the string frame Ricci scalar, as seen in Fig.~\ref{fig:Ricci}. The divergence is, however, extremely weak: it is due to corrections $\sim \log \log r$ in the asymptotic IR geometry, and therefore practically absent in any anisotropic finite temperature solution. At zero temperature for any positive $a$, though, the deep IR is expected to receive stringy corrections which are absent at vanishing $a$\@. This indicates that the limits $a \to 0$ and $\alpha' \to 0$ do not commute. Another related effect is the IR divergence of the anisotropic susceptibility for the anisotropic ground state, discussed in Sec.~\ref{sec:chia}.

We also identified several physical consequences of the novel geometry in the IR: first, is the fact that the thermal gas solutions, which represent the confined isotropic phase, is replaced by a branch of small black holes. This is similar to what happens generally in the canonical ensemble of charged black holes, where the anisotropy
parameter $a$ plays the role of the charge. This has been seen explicitly in the model used in the current paper~\cite{Alho:2013hsa}, where the resulting IR geometry was found to be $\text{AdS}_2 \times \mathbb{R}^3$, as well as in other works \cite{Chamblin:1999tk,Caceres:2015vsa,Pedraza:2018eey}. As a result, we find that linear confinement persists but only up to a certain length scale. This is evident from the behavior of the quark-antiquark potential, which shows signs of instability at large enough separations. Related to this, we observe that mesons and glueballs indeed fail to be absolutely stable but instead develop narrow widths, indicating the possibility of decaying to the $\text{AdS}_4$ vacuum.

There are some open questions and extensions of our work that are worth exploring:
\begin{itemize}
  \item \emph{Interplay with magnetic field.} A natural future extension of the present work would be to introduce a magnetic field on top of the anisotropic background and to study the interplay between magnetic and anisotropic effects. This extension has a rich parameter space as there is the possibility of introducing an angle between the magnetic field and the direction singled out by the anisotropy.
  \item \emph{Potentials and universality.} Here we used the potentials (in particular $V_g$ and $Z$) motivated by earlier work, which produce physics qualitatively similar to QCD. There is however still some freedom in choosing these potentials even taking into account all known constraints. In particular, the function $Z$ could be modified by logarithmic corrections in $\l$ in the IR, and the effects due to this could be studied. It could also be interesting to make completely different choices for the potentials and study the universality of our results. For instance, a $Z$ which grows slower in the IR (e.g. $Z \sim \mathrm{const.}$): in this case the drastic change in the IR structure between the $a=0$ and $a>0$ solutions would be absent. Studying this could therefore be interesting even though this choice seems less motivated by the comparison to QCD.
  \item \emph{String embeddings.} In the computation of the quark-antiquark potential, there is the possibility that the solutions for the string worldsheet that were found are unstable towards twisting in the bulk. Future work could study this possible instability.
      A more difficult problem to tackle is the physical significance of the unstable branches of the quark-antiquark potential. To investigate this, one would have to study time-dependent solutions of the string worldsheet. This is an inherently difficult problem.
\item \emph{Thermalization and isotropization.} Another interesting future project would be to study the time-dependence of the model more in general, and to address questions such as thermalization and isotropization of the QGP in the presence of fully backreacted quarks and anisotropic deformations (e.g. a space-dependent $\theta$ term and/or a magnetic field). As a first step, one could compute the full quasinormal mode spectrum of the system and transport coefficients. Some steps in this direction were recently taken in \cite{Itsios:2018hff}.
\item \emph{Field theory and lattice QCD.} We initiated a study of the problem directly in field theory in the Introduction, see e.g. equations (\ref{qqint3}) and (\ref{Da}). It is tempting to carry out the calculation of the quark condensate perturbatively using this setting. However, in analogy with an external magnetic field \cite{Miransky:2015ava} we do not expect the Inverse Anisotropic Catalysis phenomenon to be present at weak coupling. Nevertheless this calculation may give us hints towards a better understanding of the phenomenon. Another idea is to study the problem in an effective field theory such as the NJL, which in fact provided important insights for the IMC \cite{Preis:2012fh}. Finally, it would be very interesting to directly check our proposal on the lattice. Anisotropy on a lattice can be introduced by assigning a different number of lattice points in one spatial direction than the directions transverse to it.
This is of course not easily done as said, because introducing fermions on an anisotropic lattice turns out to be a challenging problem, see e.g. \cite{Bollweg:2018kty}. It may be easier to consider the picture given by (\ref{Da}) instead. Since this is essentially a system of Weyl semimetal with the Dirac cone separated in the left and the right parts in the momentum space by an axial gauge field, and since it is possible to study Weyl semimetals on the lattice, we expect this picture to be more suitable for the lattice studies. Yet, one has to check directly if this action is plagued by a fermion sign problem. Finally, we note that the effect of anisotropy on the critical temperature can already be checked in a pure glue setting as in \cite{Giataganas:2017koz}, which, on the lattice, is computationally easier than a model which includes fermions.

\end{itemize}
We hope to come back to these points in the near future.

\section*{Acknowledgements}

It is a pleasure to thank G.~Arias-Tamargo, D.~Giataganas, T.~Ishii, J.~Shock, and B.~Withers for helpful discussions and comments on the manuscript.  This work is
partially supported by the Netherlands Organisation for Scientific Research (NWO)
under the VIDI grant 680-47-518 and the VENI grant 680-47-456/1486, and the Delta-Institute for Theoretical Physics ($\D$-ITP), both funded by the Dutch Ministry of Education, Culture and Science (OCW).

\appendix

\section{Tachyon IR asymptotics}\label{app:tachyonIR}

In this appendix we consider the asymptotics of the full system, i.e., $x >0$, and verify in particular that the tachyon is indeed asymptotically decoupled in the IR.
In order to do this, we first assume that the asymptotic IR geometry is indeed that studied in section~\ref{sec:rolling} and study the tachyon asymptotics.
We then need to check that the tachyon diverges fast enough so that it decouples the flavor from the glue in the IR, and the obtained result is consistent.

It is useful to write the tachyon equation of motion in a different form where $A$ is used as a coordinate. Assuming $V_f(\l,\tau) = V_{f0}(\l)e^{-a_0\tau^2}$, where $a_0$ is a constant, it can be rearranged as
\be
 \frac{G}{e^{5A+W}V_{f0}(\l)}\frac{d}{dr}\left[\frac{e^{3 A+W} \kappa (\lambda) V_{f0}(\lambda) Q \tau'}{G}\right] = -2 a_0 \tau \ .
\ee
Since $\l$ evolves slowly in the IR and tachyon diverges, we may approximate $G \simeq e^{-A} \sqrt{\kappa(\l)} \tau'$. Neglecting derivatives of $\l$ and $\widetilde W$ we find
\be
 \frac{\kappa (\lambda) \tau'}{e^{5A}}\frac{d}{dr}\left[e^{3 A}   \right] \simeq -2 a_0 \tau \ .
\ee
We further insert the rough approximation $e^{-A} \simeq r/q$ where $q$ is constant. We obtain
\begin{align}
 \frac{3 \kappa (\lambda) r \tau' }{q^2} &\simeq 2 a_0 \tau \ .
\end{align}
The solution in the simplest approximation is therefore
\be
 \tau \sim r^{2 q^2 a_0/(3 \kappa(\l))}\ .
\ee
For the IR asymptotics of section~\ref{sec:rolling}, $q^2 V_g \sim \mathrm{const}$. Moreover
we have chosen the potentials such that $V_g \kappa \sim \mathrm{const}$ at large $\l$, we find that $q^2/(\kappa(\l)) \sim \mathrm{const}$ and the tachyon therefore obeys a power law in the IR. This is enough for the tachyon to decouple and for the assumptions we made above to be valid. Notice however that it may be enough to modify the subleading logarithmic corrections to the IR asymptotics of, say, $\kappa$ to change this conclusion.

\section{Holographic renormalization}\label{app:renormalization}

In this appendix we write down the counterterms needed to regularize the free energy and the anisotropic susceptibility $\chi_a$. In the case of flat boundary metric, the required counterterms are~\cite{Papadimitriou:2011qb}
\be
 S_\mathrm{ct} = - M^3 N_c^2 \int d^4x \sqrt{\gamma} \left(U(\l)+ \frac{1}{2}\Theta(\l)\partial_i \chi \partial^i \chi +\frac{1}{4} c(\l) \left(\partial_i \chi \partial^i\chi\right)^2 \right)
\ee
where $\gamma$ is the boundary metric and  we also used the fact that the only field depending on the spatial coordinates is $\chi$. The various functions are defined as follows. First, $U(\l)$ is the superpotential. Up to a choice of scheme, we can choose any solution of the superpotential equation (Eq. (1.6) in~\cite{Gursoy:2007cb}). In this work however we only use explicit counterterms to renormalize $\chi_a$ (while the free energy is obtained by integrating the first law of thermodynamics) and its renormalization is independent of $U(\l)$. The function $\Theta(\l)$ can be found in general as an integral defined in terms of the superpotential and the function $Z(\l)$~\cite{Papadimitriou:2011qb}. The function $c(\l)$ cancels a remaining logarithmic divergence and only its expansion in the UV is needed. For our purposes it is sufficient to note that the following counterterms for the susceptibility are equivalent to the general prescription:
\be
 \chi_{a,\,\mathrm{ct}} = \frac{1}{a\, V_4} \frac{\partial S_\mathrm{ct}}{\partial a} = M^3 N_c^2 \left[\int_\epsilon^{r_0} dr\, e^{3 \tilde A} Z(\tilde \l) - a^2 \tilde c\right]
\ee
where the integral arises from $\Theta(\l)$, $\tilde A$ and $\tilde \l$ give the solution at $a=0$ and $T=0$, we introduced a UV cutoff $\epsilon$, and
\be \label{ctildedef}
 \ell^{-3}\tilde c =\left.\frac{4  Z_0^2}{27   V_1\,\tilde \lambda} + \frac{Z_0^2  \left(25 V_1^2+64 V_2\right)}{216 V_1^2}\log (\tilde \lambda )\right|_{r=\epsilon} \ .
\ee
Here the UV coefficients are defined via
\be
 Z = Z_0 + \mathcal{O}\left(\l^4\right) \, \qquad V_g(\l) - x V_{f0}(\l) =  \frac{12}{\ell^2}\left(1 + V_1 \l +V_2 \l^2 + \mathcal{O}\left(\l^3\right)\right) \ .
\ee
The choice of $r_0$ and the (absence of the) constant term in~\eqref{ctildedef} reflect the scheme dependence. For the numerical evaluation of $\chi_a$ we chose $r_0$ such that $\tilde A(r_0) = 0$ in units where $\ell =1$.

\section{Fluctuation equations}\label{app:flucts}

We give here the fluctuation equations for two additional modes apart from the helicity two glueballs discussed in the main text. We have checked numerically that the spectrum of these sectors behaves qualitatively similarly as the helicity two glueballs.

\subsection{Flavor nonsinglet mesons}

The flavor nonsinglet mesons (states in the adjoint representation of the unbroken $SU(N_f)_V$) are decoupled from the glueballs. The spatial asymmetry is mediated to the meson sector (only) by the metric, and therefore the changes with respect to the isotropic case~\cite{Arean:2012mq,Arean:2013tja} are minor. In particular, as it turns out, the asymmetry does not lead to the mixing of any of the helicity zero states in our setup.

For example, the fluctuation equation for the rho mesons is
\be
\frac{1}{V_f(\l,\t)\,w(\l,\t)^2\, e^{A\pm W}\,G}
\partial_r \left( V_f(\l,\t)\, w(\l,\t)^2 \,e^{A\pm W}\,
G^{-1}\, \partial_r \psi_V \right)
+\left(\omega^2 -q^2e^{-2W}\right) \psi_V  = 0 \,
\label{vectoreom}
\ee
with plus (minus) signs in the exponents for helicity one (zero) states. Here $\psi_V(r)$ is the radial wave function for the fluctuations.

\subsection{Helicity one glueballs}

The helicity one states also turn out to be simple. For $B=0$ there are no background gauge fields, and since the action is quadratic in gauge fields, gauge field fluctuations decouple from the metric. The metric fluctuations, in the gauge where $\delta g_{\mu r} = 0$, are $\delta g_{it}$ and $\delta g_{i3}$ with $i=1,2$. Constraints arising from the $ir$ components of the Einstein equations eliminate two of these. Moreover the states with positive and negative helicities decouple. Therefore the remaining two physical fluctuations are decoupled and satisfy the same equations. Defining (considering the positive helicity for example)
\be
 \delta g_{13} + i \delta g_{23} = e^{2 A(r)} e^{- i \omega t+i q x_3} h_3(r) \ ,\qquad   \delta g_{1t} + i \delta g_{2t} = e^{2 A(r)} e^{- i \omega t+i q x_3} h_t(r) \ ,
\ee
and choosing the invariant combination $\zeta(r) = \omega h_3(r) + q  h_t(r)$, the fluctuation equation reads
\be
\zeta ''(r) +\left(3 A'(r)-\frac{\omega^2+q ^2 e^{-2 W(r)}}{\omega^2-q ^2 e^{-2 W(r)}}W'(r)\right)\zeta '(r) +\left(\omega ^2-q^2 e^{-2 W(r)}\right) \zeta (r)  = 0 \ .
\ee
At $q=0$ we see that the only change with respect to the equation for the helicity two glueballs is the opposite sign in the term $\propto\, W'(r)$.

\bibliographystyle{ucsd}
\bibliography{refs}

\end{document}